\documentclass[aps,pra,reprint,showpacs,nofootinbib,superscriptaddress]{revtex4-1}
\usepackage{amsfonts}
\usepackage{amsmath,amssymb}
\usepackage{tikz}
\usepackage{graphicx}
\usepackage{blkarray}
\usepackage{float}
\usepackage{ltablex}
\usepackage{tabularx}
\usepackage{booktabs}

\newcolumntype{Y}{>{\centering\arraybackslash}X}
\keepXColumns
\usepackage{longtable}
\usepackage{hyperref}
\usepackage{array}
\usepackage{enumitem}
\usepackage{bm}
\usepackage{xcolor}
\usepackage{graphicx}
\usepackage[utf8]{inputenc}

\hypersetup{
     colorlinks   = true,
     citecolor    = blue
}

\def\be{\begin{equation}}
\def\ee{\end{equation}}
\def\bea{\begin{eqnarray}}
\def\eea{\end{eqnarray}}
\begin{document}
\title{Quantum Key Distribution for Critical Infrastructures:
Towards Cyber Physical Security for Hydropower and Dams}

\author{Adrien Green}
\email{agreen91@vols.utk.edu}
\affiliation{Department of Physics and Astronomy, The University of Tennessee, Knoxville, Tennessee 37996-1200, USA}
\author{Jeremy Lawrence}
\email{JLawrence@epri.com}
\affiliation{Electric Power Research Institute, 1300 W. WT Harris Blvd., Charlotte, NC 28262, USA}
\author{George Siopsis}
\email{siopsis@tennessee.edu}
\affiliation{Department of Physics and Astronomy, The University of Tennessee, Knoxville, Tennessee 37996-1200, USA}
\author{Nicholas Peters}
\email{petersna@ornl.gov}
\affiliation{Quantum Information Science Section, Computational Sciences and Engineering Division, Oak Ridge National Laboratory, Oak Ridge TN 37831, USA}
\author{Ali Passian}
\email{passianan@ornl.gov}
\affiliation{Quantum Information Science Section, Computational Sciences and Engineering Division, Oak Ridge National Laboratory, Oak Ridge TN 37831, USA}



\begin{abstract}
Hydropower facilities are often remotely monitored or controlled from a centralized remote-control room. Additionally, major component manufacturers monitor the performance of installed components. While these communications enable efficiencies and increased reliability, they also expand the cyber-attack surface. Communications may use the internet to remote control a facility’s control systems, or it may involve sending control commands over a network from a control room to a machine. The content could be encrypted and decrypted using a public key to protect the communicated information. These cryptographic encoding and decoding schemes have been shown to be vulnerable, a situation which is being exacerbated as more advances are made in computer technologies such as quantum computing. In contrast, Quantum key distribution (QKD) is not based upon a computational problem, and offers an alternative to conventional public-key cryptography. Although the underlying mechanism of QKD ensures that any attempt by an adversary to observe the quantum part of the protocol will result in a detectable signature as an increased error rate, potentially even preventing key generation, it serves as a warning for further investigation. When the error rate is low enough and enough photons have been detected, a shared private key can be generated known only to the sender and receiver. We describe how this novel technology and its several modalities could benefit the critical infrastructures of dams or hydropower facilities. The presented discussions may be viewed as a precursor to a quantum cybersecurity roadmap for the identification of relevant threats and mitigation.
\end{abstract}

\maketitle

\onecolumngrid

\section{Introduction}
\begin{figure}[htp]
    \centering
    \includegraphics[width=\textwidth]{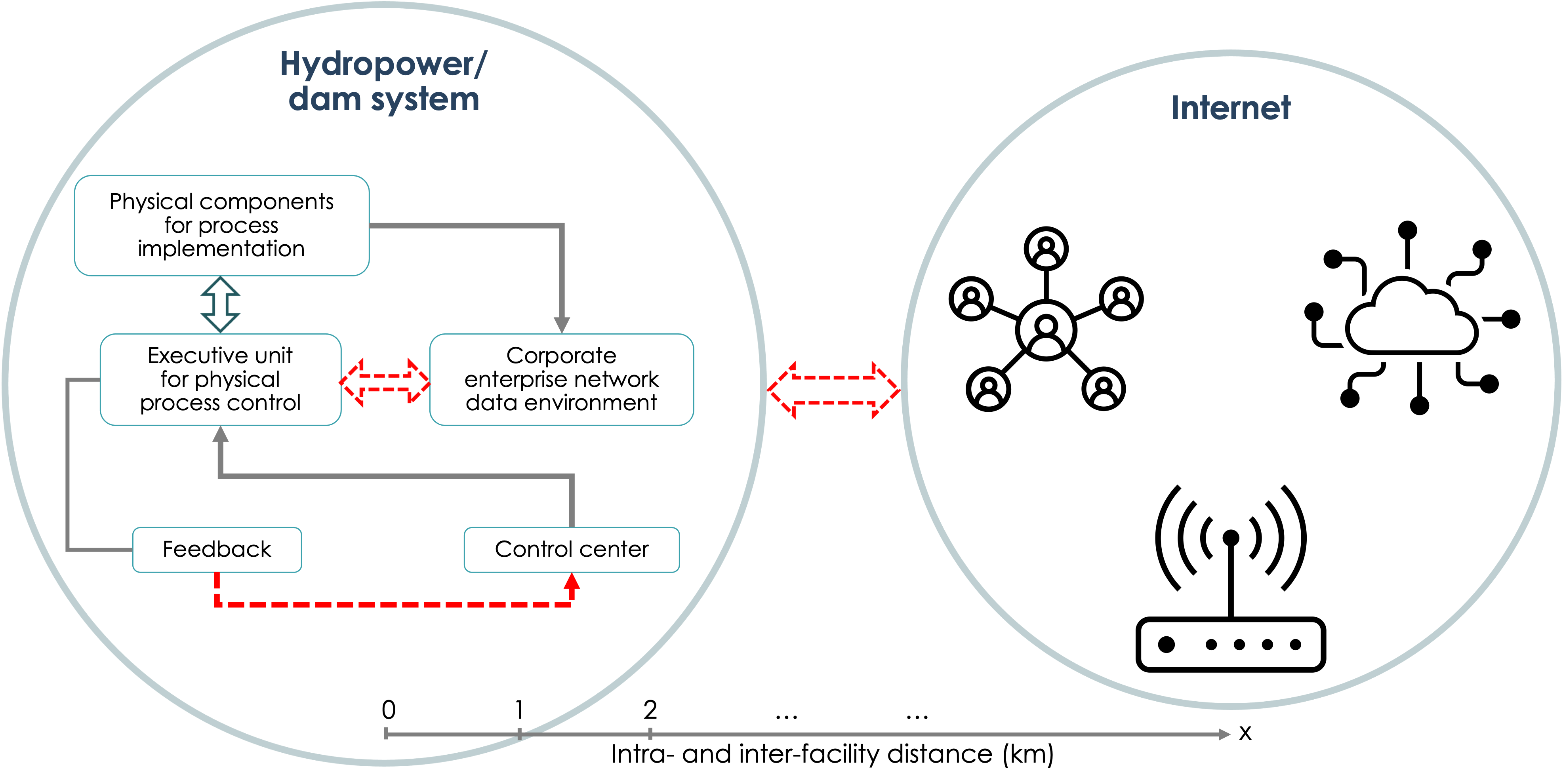}
    \caption{It is widely recognized that the existing hydro infrastructure has cyber-security weaknesses that can be exploited at both intranet (left) and internet (right) levels. As exemplified by the dashed arrows, cyber problems are ultimately due to a lack of secure communications and the presence of side channels among the various components/devices inside and outside the system.}
    \label{fig:overview}
\end{figure}

Security of critical infrastructures poses a complex and dynamic problem teeming with loopholes, weak links, and outdated measures that create an array of cyber vulnerabilities and safety concerns~\cite{rass2020cyber,whyatt2021,cisa2019}. Innovative solutions are needed to protect existing and developing infrastructure (see Rass et al.~\cite{rass2020cyber} for what constitutes a ``critical infrastructure'' and related discussions). Currently, in the US alone, less than 3 \% of the 80,000 dams produce power. Efforts to generate more clean power from these existing dams mean the utilization of advanced technologies and modernization. Therefore, digital technologies are expected to continue to be integrated with hydroelectric projects (including fleet modernization). The gain (e.g., in the efficiency from turbines and generators) that comes with digitalization and the use of advanced information and communication technologies benefit the missions and objectives of an increasing number of stakeholders in hydro energy. These efforts mean increased connectivity (e.g., enhanced remote control and monitoring of the operational conditions of the assets). Higher connectivity is also expected from optimization efforts to operate neighboring hydropower facilities across whole river systems. Predictive and intelligent maintenance~\cite{singh2020}, higher efficiency operation, development of digital twins, etc., all require communication of measurement results and associated data analysis from many components and equipment, often in real-time. Higher connectivity, that is, a larger number of communications channels, means a larger cyber-attack surface, and consequently, more risks, as depicted in Fig.~\ref{fig:overview}. A brief summary of some of the basic security issues is provided in Table~\ref{table}, see also relevant discussions by Ratnam et al.~\cite{RATNAM2020106833}. In what follows, for convenience, some relevant terms invoked are defined in Table~\ref{table:glossary}. 

Clearly, the noted risks associated with exploiting the weaknesses of communications channels need to be addressed. However, known classical (non-quantum) encryption techniques cannot eliminate such risks (for a simple classical encryption example, see Appendix~\ref{appendix:examples}). This is because, to protect the confidentiality of the communicated messages, classical security utilizes the mathematical complexity of classical cryptography techniques (which ultimately can be figured out with sufficient computer power), as opposed to quantum approaches which capitalize on fundamental physical laws. These laws mean that any attempts to intercept or read off the information will disturb the fragile quantum states carrying the information. The imposed fundamental limits here mean that there is no amount of care one can exercise that would enable this process without creating a detectable quantum disturbance. Thus, without embracing quantum solutions, the long-term security of hydropower and dam infrastructure,  remains uncertain.  

Traditional encryption is currently used to validate the legitimacy and authenticity of the sender and receiver, while also obfuscating the information from an attacker. This means that even if communication 
is intercepted~\cite{chen2011lessons}, it cannot be read or understood unless the attacker has the decryption key. Traditional methods rely on secure creation and exchanges of keys to ensure end-to-end protection. 
Current attacks against encryption include incorrect implementation of encryption in software (vulnerabilities), attacks against the users, supply chain attacks, compromise of the keys, brute forcing the message, or analyzing the encrypted communication to derive the key. With the future advancement of quantum computing, the speed at which brute force attacks can successfully decrypt communications (using many current algorithms) could render them insufficient. 
The problem of infrastructure vulnerability means that attempts to evade security measures including cyber, malware, and side-channel attacks may generate results~\cite{rass2020cyber}. Reported attacks on dams and other critical infrastructure have revealed significant cyber-security gaps and problems in existing infrastructure, which is ultimately due to a lack of secure communications channels. Physically, the two primary channels over which information, commands, and instructions are conveyed/exchanged are either optical fiber or free space with no other comparable alternatives. Both these channels can be exploited by attackers to threaten the assets. This article proposes a solution and how it may be applied to this problem. 
Clearly, any solution necessarily should be commensurate with the rapidly growing and diversifying information and communications technologies encompassing edge computing and sensing~\cite{passian2019,farahi2012}, IIoT (the Industrial Internet of Things), IoE (the Internet of Energy), etc. Such a solution must position the hydro security infrastructure for resiliency against increasingly advanced and sophisticated attacks. 

Emerging quantum technologies promise to solve the security problem of communications channels. The most well-established quantum communication technology is currently quantum key distribution (QKD), which has been shown to achieve information-theoretic security (ITS), meaning it does not rely on any technology assumptions, such as what problems are difficult to compute. Such a solution has already been demonstrated in the form of the deployment of state-of-the-art QKD-based communications technologies across the electric grid~\cite{Alshowkan2022,evans2021,grice2013,kuruganti2014}. To date, as technology transitions from research labs to the commercial sector, only a few commercial QKD systems have made their way to the market. These commercial systems are of basic design and not yet fully adaptable to the hydro environment. An evaluation and comparison of all QKD modalities against hydropower system’s requirements is 
needed (see Table~\ref{tab:qkd_comparison}). Logically, one may categorize noise sources in QKD operating in a hydropower environment into two main categories: those induced by the environment, as listed in Table~\ref{noise}, and those independent of it, as listed in Table~\ref{intrinsic:noise}, with related discussions elsewhere~\cite{scarani2009security,gobby2004quantum,yuan2007high,rosenberg2005practical,hiskett2007long}.
While noise sources independent of the environment may be addressed with advances in technology and improved equipment, those induced by the dam environment may require specialized solutions tailored to the unique challenges posed by such a setting.
Separating these categories could help in better understanding and mitigating the noise sources.

\section{Objective}

\begin{figure}[htp]
    \centering
    \includegraphics[width=\textwidth]{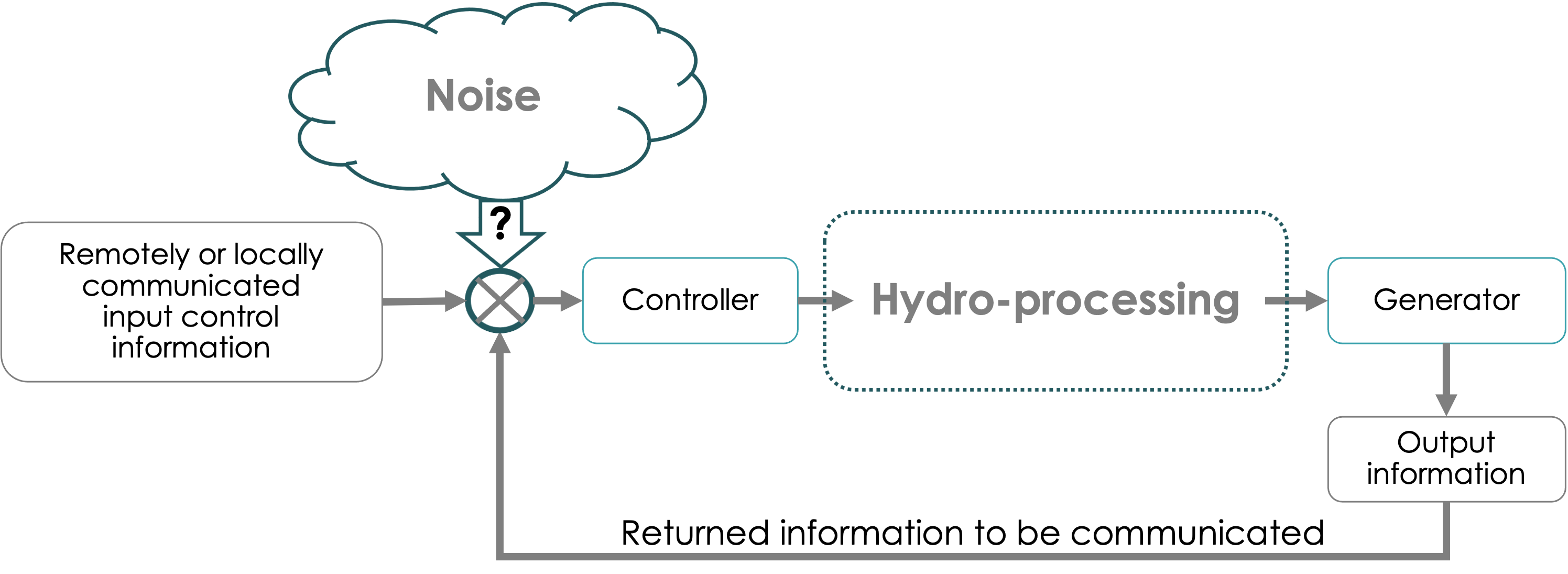}
    \caption{Schematic depiction of feedback loop for control of the hydro-electric process. Multiple points of vulnerabilities can be readily identified.}
    \label{fig:example0}
\end{figure}

\begin{figure}
    \centering
    \includegraphics[width=\textwidth]{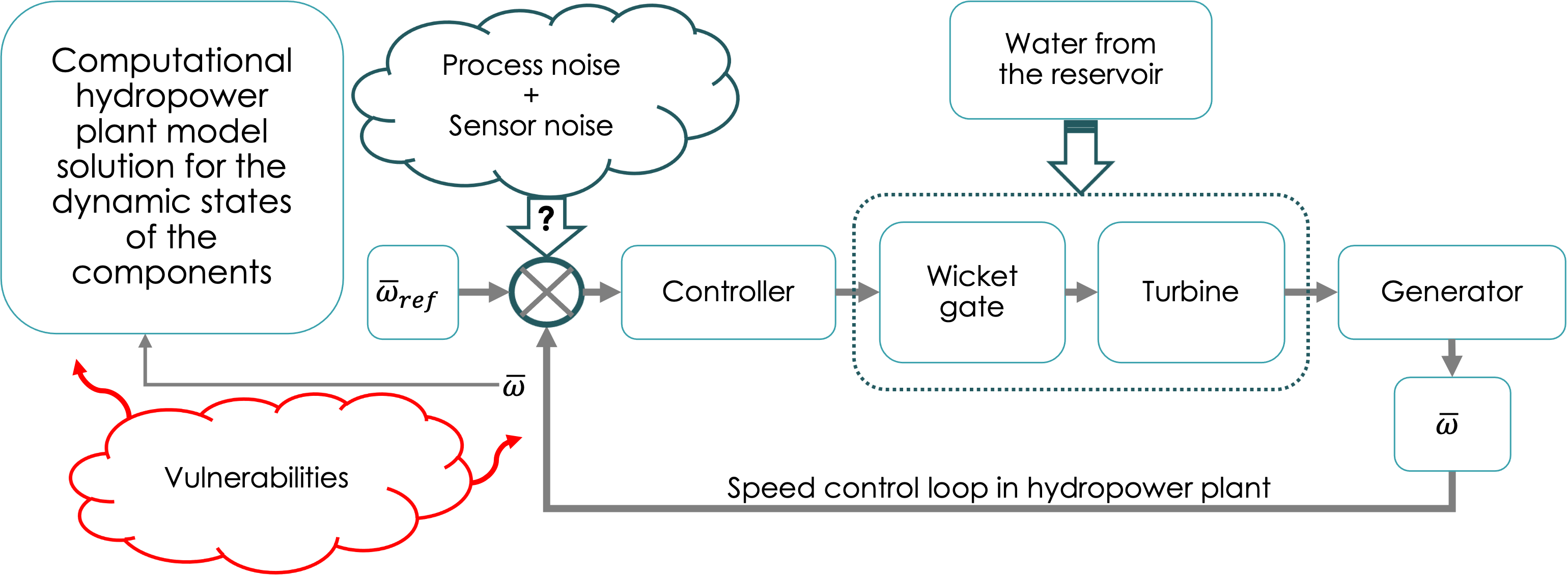}
    \caption{An example of side-channel threats in a hydropower plant. Secure speed control of the generator is critical for safe operation. During remote control, cyber-attacks may alter the commands leading to altered input values for the speed control of the generator. The displayed differential equations (see Chandra et al.~\cite{chandra2021}) for the time dependence of the states including frequency change $\Delta\bar{\omega}$, turbine water velocity $\bar{u}_t$, gate opening $\bar{g}$ and the pilot actuator position variation $\Delta\bar{x}_e$ serve to emphasize the many parameter and functional dependencies of the hydropower plant model. Possible side-channel attacks may take the form of altering the signal returned by the feedback loop. These communications channels can be made secure using a future implementation of internet-level, intranet-level, and even “device-level” QKD.}
    \label{fig:example}
\end{figure}

Our objective is to elucidate the utility of QKD for protecting hydropower assets and articulate what quantum security technologies can bring into the critical hydro infrastructure security domain. As simplistically depicted in Fig.~\ref{fig:example0}, a hydropower system is composed of many networked sensors, control systems, and operators which need to communicate with each other over geographically diverse locations. A relevant use case of secure communications may therefore be the implementation of a communications channel between the control room and an equipment controller. This would not only prevent cyber-attacks but also any side-channel attacks. 
The communications link corresponding to this connectivity can be made secure with long-term security provided by QKD. Any information to be exchanged between a party on the internet and a party on the dam network requires encryption or authentication. QKD can share unique private keys that can be used for this purpose (for a simple quantum encryption example, see Appendix~\ref{appendix:examples}). Therefore, the generation of secure keys over the communications link is the first step (see Fig.~\ref{fig:example}).
Reaching a working understanding of the technical layout specific to a hydro facility could begin by building on previous work in carrying out cyber technical risk assessments including that of hydropower systems and dams~\cite{evans2021,lawrence2020}, and development of holistic cybersecurity risk reduction framework for fossil generation facilities~\cite{lawrence2020}, as well as deployment of quantum communication for grid security~\cite{kuruganti2014} and~\cite{evans2021,grice2013}. 
Given that there are many such plant and systems-level examples of a cyber threat to physical devices of the power plant, during any pilot project, identifying a reasonable location in a dam or an equivalent testbed to implement the QKD is prudent. This important step lays the foundation for developing a similar use-case development methodology that can be applied across various hydro facilities. One may envision research and development of a security use-case taxonomy that can serve the broader energy infrastructure landscape. Such a taxonomy should be of direct benefit to the stakeholders since planning, marketing, energy distribution, customer privacy, service quality, and many other aspects of energy economics can be impacted by a better understanding of the security risks involved.

QKD is currently at the forefront of innovative communications technologies and is typically advertised as the next-generation security technology that, unlike conventional techniques, does not expire when bigger computers are built. QKD has been demonstrated for securing communications between parties on Earth~\cite{wang2022} as well as between Earth and satellites~\cite{liao2017,yin2020}. To implement QKD, we first note that there are different modalities of QKD. Despite, some modalities still being subject to intense research, new results are leaving research labs, and commercial systems are entering the market. For a brief description of the QKD modalities versus the specific requirements of the present use case see Table~\ref{tab:qkd_comparison} (see also the recent survey by Sharma et al.~\cite{sharma9520678}

\section{Quantum key distribution and its implementation}
In 1984 the BB84 protocol was introduced by Bennett and Brassard~\cite{scarani2009}. Today, BB84 is just one instance of many possible QKD implementations that rely on the laws of quantum mechanics to share secret keys. As an example of a standard implementation, the polarization of individual photons can be used to encode information. In BB84, the simplest, oldest, and most developed protocol, Alice sends a 
signal of single photons with random polarizations to Bob, and since quantum states are disturbed when measured, any noise is attributed to an eavesdropper resulting in quantum bit error ratio (QBER), defined as:
\begin{equation}
    \text{QBER}=\frac{\text{number of error bits}}{\text{total number of bits exchanged}}. 
\end{equation}
If the QBER is low enough, the two parties can distill a secret key. For many QKD protocols, the process to distill a secure key involves four main steps: raw key exchange, sifting, error correction, and privacy amplification, as shown in Fig.~\ref{fig:example}. Specifically, in the BB84 protocol, after the raw key is shared through the quantum channel, the sifted key is generated by Bob announcing, over a public classical channel, the basis he used to measure each photon. Alice then compares this with her basis choices. The bits corresponding to mismatched bases are discarded, resulting in the sifted key. Any QKD protocol will have errors in the sifted key due to the experimental imperfections and potential eavesdropping, which are corrected using an algorithm over the authenticated public channel. Finally, privacy amplification is also performed over the public channel to minimize the potential information about the key that Eve may have gathered in the previous three steps. Generally, QKD can be divided into discrete variable (DV) and continuous variable (CV) implementations. DV encoding, such as the polarization example above, involves detecting single quantum states with direct detection single-photon detectors, and results in discrete measurement data. In continuous variable encoding, homodyne detection is used to measure continuous variables of quantum light, namely phase, and amplitude, which carry information between Alice and Bob. CV QKD protocols generally have the potential for higher key generation rates than DV protocols, especially in the presence of lossy channels. This is due to the encoding and detection methods employed by CV systems. However, CV protocols, especially in high-loss scenarios, face challenges related to error reconciliation, given the Gaussian noise characteristics of their keys~\cite{scarani2009}. A benefit of CV protocols is that they can be implemented in shared fiber with classical communication systems without the destruction of the quantum signal~\cite{qi2010}. QKD can also be divided into entanglement-based and prepare-and-measure implementations. In many entanglement-based QKD protocols, entangled states of light are generated and shared between Alice and Bob. This can be done by a third party or by one of the participants, such as Alice, who then sends one of the entangled particles to Bob. On the other hand, prepare-and-measure protocols, like BB84, involve Alice preparing a quantum state and sending it directly to Bob. For distances relevant to hydropower dams, prepare-and-measure protocols are the most practical and achieve the highest speeds~\cite{Scherer2011}. In practical QKD implementations, information is often encoded into weak coherent pulses, such as those from faint lasers. While single-photon or entangled photon sources are ideal for QKD, their practical implementation can be challenging. The decoy state protocol allows weak coherent sources to be used effectively by mitigating certain eavesdropping threats, making it possible to utilize cheaper and faster equipment for quantum light sources without compromising security~\cite{Lo2005}. This encoding method has evolved to what is called decoy state QKD because certain laser pulses will act as decoys to an eavesdropper for security purposes. Given that decoy-state BB84 is among the most developed protocols and gives the highest data rates for the relevant distances, it makes for a suitable field trial. 
\begin{figure}[htp]
    \centering
    \includegraphics[width=\textwidth]{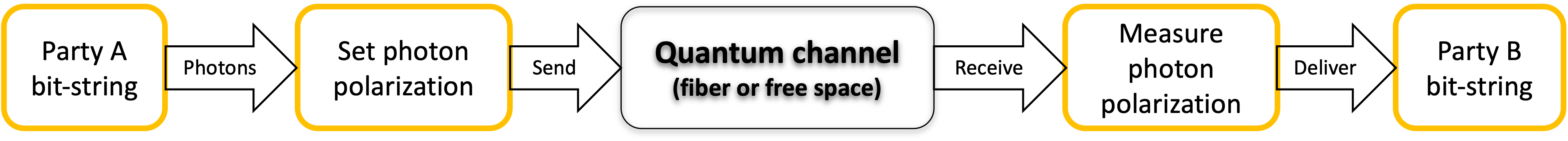}
    \caption{Schematic depiction of QKD. Two parties generate random bit-strings that are encoded in light pulses. The pulses are sent off via fiber or free space to a receiving party, which is measured and  converted to bit-strings. A message may be encoded via variations in the polarization of the light, as shown in the second box from left, or in its phase (see R. Wolf~\cite{ramona:book} for a formal introduction to these operations).}
    \label{fig:workflow}
\end{figure}

Studies of scientific and technical issues surrounding the security  of practical implementation of QKD have illuminated the possibility of various conceivable side-channel attacks. Realistic expectations from the performance of QKD subsystems mean that QKD must be carefully implemented~\cite{lutkenhaus2000security,makarov2006optical} to avoid for example Trojan horse and photon-splitting attacks. These attack scenarios are difficult to mount on practical QKD systems which are evolving to safeguard against these and future loopholes  in QKD implementation. 
The implementation of QKD begins by building its physical arrangement, which is composed of light sources, optical components to manipulate light, detectors, and measurement and processing 
electronics. Alternatively, one may explore the use of commercial QKD systems to generate the needed keys for information encoding. The information encoded by doing a bit-by-bit exclusive “OR” with these keys will be secure if the key is kept private, and the key is larger than the message size. Successful implementation of QKD is measured by generating keys using the physical realization of the diagram shown in Fig.~\ref{fig:workflow}.
\begin{figure}[htp]
    \centering
    \includegraphics[width=\textwidth]{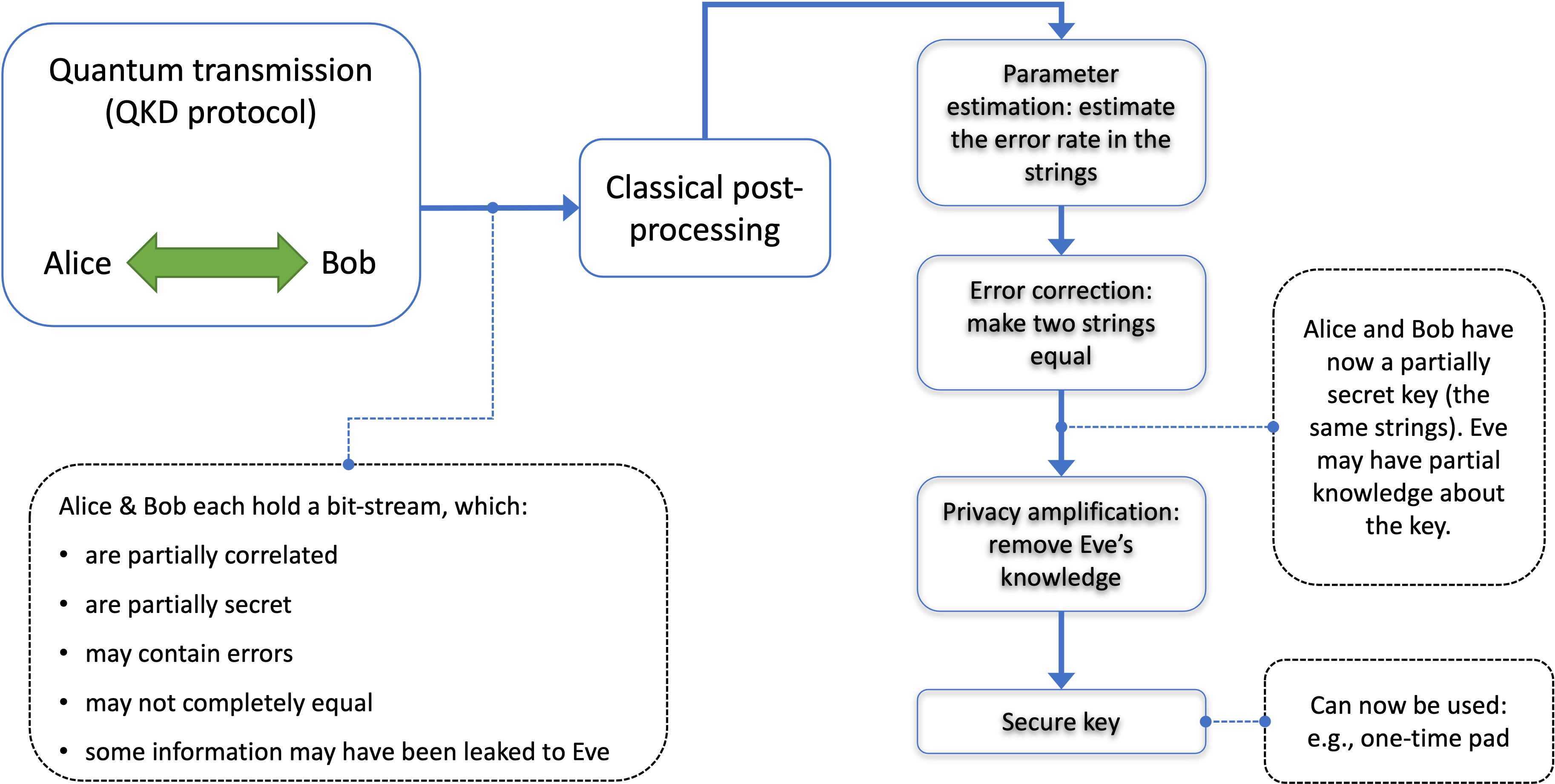}
    \caption{In the basic form of QKD, the two communicating parties, Alice and Bob carry out quantum optical operations to prepare some bit-strings. After quantum operations are completed, post-processing is required. These include applications of various mathematical and numerical processing to generate the final keys. If an eavesdropper (Eve) attempts to intercept the processes, laws of quantum physics reveal such attempts as errors in the quantum signals. See Table~\ref{table:postp} for further description and R. Wolf~\cite{ramona:book} for details.}
    \label{fig:protocol}
\end{figure}

The key questions to be answered are where QKD can be deployed in a hydropower communications network, and how it can be integrated with existing command and control interfaces. Of importance is the frequency of communications and the requirements of QKD-based one-time-pad (OTP) approaches to meet this need, in addition to communications latency requirements.
A summary that systematically and collectively presents the application of QKD to the hydroelectric domain will help to lead the wider utilization of quantum security in hydro infrastructures. Our goal is stimulate the production of this document, which would also complement those focused on classical cybersecurity.

\section{Approach}
An approach to the security of hydropower assets based on quantum technologies should naturally be relevant to other energy infrastructure security science and technology. Therefore, a high degree of connectivity is expected amongst several works in the energy research portfolio. For example, in fossil power plant cyber security, some previous investigations focused on: 1) assessing how cyber risk changes across a facility’s life cycles, 2) performing consequence analysis to prioritize high-consequence events, 3) identifying the digital asset attack surface in sensors, instrumentation, and control equipment, and 4) mitigating cybersecurity control- or countermeasures~\cite{lawrence2020}. Such reports~\cite{lawrence2020}, describe the current industry 
cybersecurity best practices in fossil generation that are based on the first principles for cybersecurity engineering. Another specific example is related to grid security, where previous work focused on: 
\begin{itemize}
\item Conducting an analysis of commercial QKD capabilities~\cite{Alshowkan2022,Anton_review}.
\item Conducting an analysis of smart grid security needs~\cite{moreno2021comprehensive}.
\item Identifying the highest value security needs that can be met by QKD~\cite{kong2020review}.
\end{itemize}
The performed analysis was based on the NIST Framework and Roadmap for Smart Grid Interoperability Standards~\cite{kuruganti2014}, as described in the related report~\cite{gopstein2021}. To achieve the technical objectives above, we may consider the following discussions. It is important to note that operational technology (OT) architectures used in hydropower control and safety systems present unique challenges and considerations from standard information technology (IT) deployment of QKD. OT systems rely on legacy equipment, proprietary, and unique operating systems, specialized protocols, and unique architecture requirements. In addition to the end-use case identification for QKD in hydropower, it is important to consider the impact of these architectures and infrastructure on QKD.

\subsection{Use-case definition for quantum security in hydro}
In creating an optimum use case, previous experience, e.g., in performing a cyber security risk assessment of other architectures, may be leveraged to show the holistic security benefit of the QKD solution. Use cases anticipated to be identified include remote monitoring and control, remote sensor, and IIOT deployment. Critical communications which rely on strong authentication in hydropower include:
\begin{itemize}
\item	Securing remote interactive access (control, maintenance, and repairs)
\item	Remote monitoring (remote sensors for control/safety/monitoring, remote monitoring only centers with unidirectional traffic)
\item	Vendor monitoring
\item	Supply chain security (validation of the authenticity of software and supply chain communications)
\end{itemize}

Control systems and operational technology (OT) rely on specific protocols for communications between field devices, programmable logic controllers, management servers and workstations, and other control system components. Many hydropower facilities that were designed with SCADAs (supervisory control and data acquisition systems)~\cite{ceser2011,Alrefaei2022AnOO} are being upgraded to distributed control systems (DCS) as hydropower facilities are undergoing component and digital modernization. The control systems must be carefully architected to provide reliability and safety. Latency and reliability of the communications are crucial in these applications and should be considered. A priority use case should:
\begin{itemize}
\item	document how it capitalizes on the specific environment of the dam/hydro facility or hydro testbed from a security point of view,
\item	Identify security benefits/disadvantages of the QKD relative to traditional methods, in the identified use cases, 
\item	document how it highlights the practical (logistical) suitability/applicability of the QKD for implementation within the dam/hydro environment/testbed,
\item	document a reference architecture for deployment in the selected use case,
\item	identify operational impacts on QKD deployment,
\item	highlight how the use case contributes to the missions of hydropower research facilities.
\end{itemize}

\subsection{Integration of the QKD system with the hydro communications system.}
QKD is a novel quantum-based cybersecurity tool that allows for the generation AND secure distribution of truly random number streams. Field demonstration of  QKD has been reported in the case of a real-world electric utility optical fiber network~\cite{evans2021}.
A “key” is simply a string of bits, that is, a sequence of 0s and 1s, and a “message” is in the form of a bit string. The end goal here is the successful use of keys generated using QKD by the communicating parties. For example, when the two communicating parties, share a private key, they can use that key to encrypt any messages they intend to send and decrypt any messages they receive. This encryption prevents eavesdropping from accessing any information in the messages. This could for example take the form of the complete set of communications needed for remote control of a dam, or communication for a SCADA system~\cite{ceser2011}. This would ultimately entail generating random bits, that are supplied to a computer hard drive or memory, at two (or more locations). These bits are then to be used for the encryption of the messages between the two locations. 
The fiber-based QKD is highly versatile as fibers are immune to electromagnetic interference, and mechanically flexible so that they can penetrate confined areas, elaborate machines, and devices. 
The QKD process begins with a quantum transmitter (typically referred to as Alice, as indicated in Fig.~\ref{fig:protocol}). The sender will have to generate light and prepare it in a specific quantum state. These light pulses, representing bit-strings, are then sent into an optical fiber to travel to another location, where they can be detected by a quantum receiver (typically referred to as Bob, as indicated in Fig.~\ref{fig:protocol}), at the other end of the fiber. After concluding the quantum operations between the two communicating parties, to generate the final keys, the bit-streams must be processed using post-algorithms. After processing the keys, as shown in Fig.~\ref{fig:protocol}, they can be used to protect the information between communicating entities (users, control systems, sensors, actuators, SCADAs, etc.). 

\begin{figure}[htp]
    \centering
    \includegraphics[width=12cm]{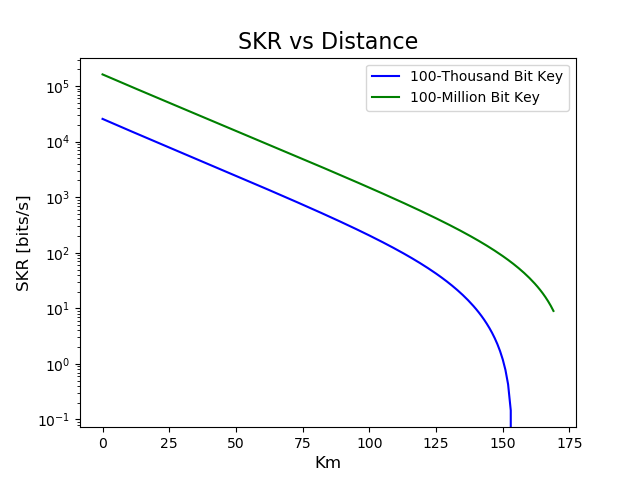}
    \caption{ The secure key rate as a function of the distance between communicating parties, derived from fiber-based loss and error models according to Eqs.~\ref{eq1} and~\ref{eq2}. This representation assumes a decoy-state BB84 variant of QKD, with parameters grounded in real-world and feasible experimental setups.  As can be seen, for longer key lengths, the allowable communication distance before the SKR becomes impractical is reduced. An example of such a system, simulated at the University of Tennessee for a free-space deployment, is presented in~\cite{Moschandreou2021}. Conventional devices, including a 1550 nm continuous wave laser and single photon avalanche detectors, were employed in their work. }
    \label{fig:graph}
\end{figure}

QKD operation means that new keys are attempted to be generated.  The conditions of the link and specifics of the QKD system determine the key rate. The key bit strings will be written to a file located on both Alice and Bob's portions of the node. The keys and hydro communications are then collected on a local computer where the encryption and authentication are implemented~\cite{Alshowkan2022}. The most computationally efficient (and therefore lowest latency) method remains the one-time pad (OTP) method, where a message is combined with the exclusive OR operation (XOR) and the key. OTP exhibits ITS (information theoretical security), i.e., it is secure regardless of an adversary’s computational power, with the following requirements: (1) the keys must be truly random, be kept secret, and are used once only, and (2) the message length is less than or equal to the length of the key. The resulting communications are then sent out through a classical transceiver. An experimental demonstration of relaying keys between relevant hydro infrastructure locations could conclude after a QKD operation over a given period (e.g., $\sim$ hours). Such an experiment could implement QKD over a metro-area distance (typical of hydro facilities) using a commercial QKD system. The key metric governing QKD system performance is the secret key rate (SKR) – the (average) number of secret bits generated and distributed securely between parties per second. SKR, while largely dependent on the type of system and QKD protocol employed, is ultimately determined by the optical loss on a given fiber link. This loss $\gamma$, expressed in units of dB, is largely due to the fiber's attenuation $a$~dB/km, which typically arises due to absorption and scattering mechanisms and can be written for a fiber of length $L$~km as: $\gamma = aL$.
 It is crucial to minimize the losses, which also can be exacerbated by fiber-to-fiber connectors, sharp fiber bending, and splicing. High losses can compromise the effectiveness and security of the QKD process. The greater the optical loss, the lower the SKR, and vice versa. In situations where the optical link loss is significant, the SKR can be zero, indicating that no secret keys can be generated. From a practical standpoint, as noted above optical losses receive contributions from two main factors: the physical distance along the fiber between two points (length attenuation); and splice, or connection loss. The former is indicative of intrinsic material losses in the optical fiber itself. Modern optical fibers exhibit $\approx (0.2-0.5)$~dB/km depending on fiber type, manufacturer, and wavelength of light. The latter is indicative of fiber-to-fiber connections, including in-field splices during deployment or following line breakage, and patch cable connections within a communications facility or substation. For this attenuation range, if, as simulated in Fig.~\ref{fig:graph}, the fiber is 175~km long, the total loss in the fiber will be in the range: $\gamma = aL=(34.4-87.5)$~dB.
 Consequently, a viable QKD deployment must evaluate the optical fiber conditions from points A to B, and link-specific SKR must be measured. 

In QKD, the eventual length of the secure key is determined by several factors including channel noise, error rates, and the specifics of the chosen protocol. While longer data collection times can yield larger secret keys, this could introduce delays before the key becomes available for encryption purposes. This is due to the need for post-processing steps like error correction, privacy amplification, and particularly the estimation of parameters such as the quantum bit error rate (QBER) using a substantial portion of the raw key. For real-world applications in hydropower plants, system optimization becomes vital. For instance, when several single-photon detectors in a command center are shared between remote links, the time each remote device utilizes a given detector should be optimized to reduce the total number of necessary detectors. This not only aids in efficient key generation but also in minimizing costs associated with hardware. The key rate or efficiency is not determined by a pre-selected length but rather emerges from the conditions of the quantum channel and post-processing. Practical QKD systems also need to address finite-size effects, where the security of the generated key can be influenced by statistical fluctuations, especially in systems with limited exchanged qubits. These effects become crucial in real-world applications such as hydropower plants, where reliable and timely key generation might be essential. An understanding of the communication frequency and topology between devices in such environments will be pivotal in tailoring QKD systems for optimal performance and cost-efficiency. 

Figure~\ref{fig:graph} depicts how the SKR varies with distance for different key lengths, highlighting the impact of channel loss on the key rate. Similarly, Figure~\ref{fig:graph} illustrates the SKR's sensitivity to misalignment angles in the system.
The secure key rate (SKR), as derived from the theoretical framework introduced by Lim \textit{et al.}~\cite{lim2014}, illustrates this dependency. In the protocol proposed by Lim \textit{et al.}, Alice sends Bob randomly polarized coherent states in two orthogonal bases: X and Z. While the X basis contributes to the secure key, the Z basis states are publicly disclosed to estimate the error rate in the X basis. The effective secure key length \(L_{\text{key}}\) is then described by:
\begin{equation}\label{eq1}
    L_{\text{key}} = s_{x,0} + s_{x,1} - s_{x,1}h(\phi_x) - \mathrm{Leak}_{EC} -6\log_2\left (\frac{21}{\epsilon_{\text{sec}}}\right) - \log_2\left (\frac{2}{\epsilon_{\text{cor}}}\right ),
\end{equation}
where \(s_{x,0}\) and \(s_{x,1}\) represent the number of dark counts and single-photon counts at Bob's detector, respectively. The term \(\phi_x\) denotes the error rate in the x basis. The binary entropy function \(h(\phi_x)\)~\cite{wilde_2017} is given by:
\begin{equation}\label{eq2}
    h(\phi_x) = -\phi_x\log_2 \phi_x - (1 - \phi_x)\log_2 (1 - \phi_x),
\end{equation}
which captures the maximum information Eve can deduce about the total key given the shared bits used to determine the error rate. As such, the term \(s_{x,1}h(\phi_x)\) must be subtracted from the total to yield a portion of the key that remains concealed from Eve. \(\mathrm{Leak}_{EC}\) encapsulates the information exposed during error correction, while the concluding terms address finite size effects. A deeper analysis, especially of terms rooted in the X basis signals and shaped by the sacrificed Z basis signals, is detailed in~\cite{lim2014}.

Figure~\ref{fig:graph} illustrates how the choice of key length, influenced by finite size statistics, affects the secure key rate and associated generation time. Specifically, at a distance of 1~km, starting with the aim to distill a 100-million-bit secure key yields a sifted key rate of approximately 98~kbps. In contrast, aiming for a 100-thousand-bit secure key results in a sifted key rate of about 25 kbps. Nevertheless, the larger initial key length necessitates a longer data collection duration: 17~minutes compared to the mere 4 seconds required for the shorter initial key length. In a continually operating secure communication system, these trade-offs highlight the importance of preemptive considerations. Factors such as communication frequency, average message size, and required security levels play a pivotal role in optimizing system performance and cost. Such metrics also influence choices regarding the QKD protocol, quantum encoding strategy, and equipment selection, ensuring that the system meets or exceeds the desired performance benchmarks.

The unique environment of a hydropower plant (see Table~\ref{noise}) introduces specific imperfections crucial in the context of a QKD system. Predominantly, additional loss and noise from such a setting can elevate the QBER rates for the QKD system. Given the finite size effect in the context of secure encryption, variations in QBER invariably gravitate towards the maximum bound of error. Hence, fluctuations introduced by the dam environment can be quite influential. The channel error model used for QKD simulations is described by Eq.~\ref{eq3}~\cite{lim2014}:    
\begin{equation}\label{eq3}
e_k = p_{dc} + e_{mis}(1-e^{-\eta_{ch}k}) + \frac{p_{ap}D_k}{2},
\end{equation}
where $e_k$ is the error rate for a coherent pulse with intensity $k$, $p_{dc}$ is the background noise rate of the detector (dark count rate), $p_{ap}$ is the after-pulse probability, and $D_k$ is the detection rate. $\eta_{ch}$ represents the loss due to the fiber optic cables and is given by $\eta_{ch} = 10^{-0.2L/10}$, with $L$ being the fiber length in km. The term $e_{mis}$ stands for the probability of error due to polarization changes in the channel and can be influenced by environmental factors.

Given that turbine operations, with their  frequencies typically around 1~Hz and 30~Hz (Table~\ref{noise}), can introduce vibrations of typically less than 1 mm in amplitude and that generators, operating at either 50 Hz or 60 Hz, induce similar amplitudes of vibrations, the environment's vibrational noise becomes crucial. Environmental factors, from seismic activities to localized events like machinery operations, can further introduce vibrational noise that influences the polarization states in fiber optics, which are sensitive to such changes~\cite{ding2017polarization}. Understanding and mitigating these effects is pivotal for QKD. For instance, correlating the vibrational frequency and amplitude data with phase changes in the QKD system could enable real-time counteraction of potential misalignments. As observed in Fig.~\ref{fig:graph}, longer raw key lengths facilitate secure key generation even at higher misalignments, an important consideration in a noise-prone dam environment. To ensure high secure key rates, hydro QKD systems should adopt polarization stabilization techniques~\cite{mekhtiev2021polarization,wu2006stable}. Typically, stabilization is achieved through feedback loops that monitor changes in the final state, enabling the sender to effectuate corrections. Given that numerous dams employ fiber-based sensors~\cite{inaudimonitoring,de2021fiber}, integrating such vibrational and noise data into the stabilization algorithm offers a promising avenue to maintain optical alignment, optimizing the QKD system's performance.
\begin{figure}[htp]
    \centering
    \includegraphics[width=12cm]{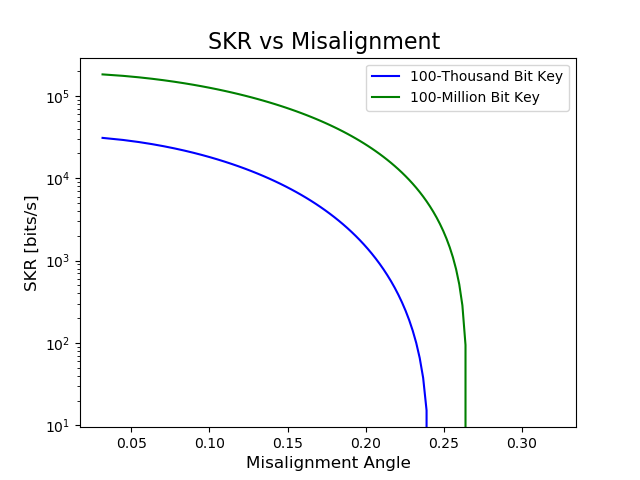}
    \caption{The secure key rate, depicted as a function of the misalignment angle, relies on Eq.~\ref{eq1} and Eq.~\ref{eq3}. Notably, commencing with a longer sifted key provides a buffer against higher degrees of polarization rotation in the quantum channel. }   
    \label{fig:graph2}
\end{figure}

With measured SKR metrics in hand, the most appropriate cybersecurity strategy for QKD-secured hydro 
communications should be evaluated. This can be guided by the following principle: if the classical communication bandwidth needs for hydro is less than the SKR, then OTP may be employed (i.e., number of QKD bits > number of classical bits requiring encryption). However, if the classical communication bandwidth needs exceed the link-specific SKR, then an ITS alternative method must be employed where the same QKD key can be used to authenticate multiple messages which can be done as long as a QRNG supplies a new nonce~\cite{Alshowkan2022}.
Finally, regardless of the cryptography option above (OTP or authentication), the interface required to supply QKD keys to the user/application must be developed. This is dependent on the type, vendor, model of the user/application, and methods by which the device allows ingestion of external (i.e., QKD) key material. This final experiment will demonstrate the encryption/decryption of realistic hydropower command/control communications using QKD-supplied keys. 
Performance challenges include SKR changes with variations in the environment in which the subsystems of QKD are to operate. Dealing with various sources of noise (including those in Table~\ref{noise}) is of great importance in the successful generation of keys. 
For example, when both quantum and classical light are considered over the channel, a concern arises from ``Raman noise’’, which is unwanted light generated in the fiber material due to the use of stronger classical light. This effect is particularly pronounced when the wavelengths of the quantum and classical signals are closely multiplexed. However, when they are in separate bands, such as the quantum signal in the O band and the classical signal in the C band, the impact of Raman scattering is substantially mitigated~\cite{Peters_2009,Chapuran_2009}.
Appropriate hardware choices can be made to better address the challenges and noise sources associated with the specific setting of the hydro facility. 
Although several QKD protocols exist, the well-established Bennet–Brassard protocol (BB84 protocol) makes for a suitable trial. Using the software, the dam communications can be interfaced with QKD keys. Similar experiments have been effectively performed to analyze and address implementation challenges facing the deployment of QKD systems in critical infrastructure, for example as demonstrated in the recent field test of three QKD systems on a real-world electric utility optical fiber network~\cite{evans2021}, where one endpoint was a hydro/dam. 

\section{Conclusions and outlook}
Witnessing the overall growth trends of quantum technologies in solving energy infrastructure problems, the presented material introduced the specific use case of the hydro energy sector. The preliminary discussions presented may help the creation of a more specialized “Quantum for Hydro” roadmap. Parameters that characterize the hydro/dam environment, as summarized in Table~\ref{noise}, are different from those found in a laboratory setting. Some of the parameters likely also differ from those encountered in the electric power grid substations where QKD has been demonstrated. Typical ambient real-world environmental conditions of importance to the performance of any technical measuring device include temperature, humidity, and various noise levels (electromagnetic, acoustic, wind, corrosion, contamination, etc.). QKD is built from sensitive optical and electronic components and devices, each with a set of specifications. Therefore, these parameters, if for example out of range, could impact the rate of key generation. Consideration for application of quantum sensing for environmental monitoring may also prove useful in conjunction with QKD. In compiling such a roadmap, consideration of important issues such as interoperability between QKD systems that operate with dissimilar implementations must be considered. In doing so, QKD standards by the ETSI Quantum-Safe Cryptography Working Group, and QKD network and QKD systems activities within ITU-T SG13 and SG17, respectively, will be put in perspective~\cite{evans2021}. 

\acknowledgments
Funding for this work was provided in part by the U.S. Department of Energy (DOE), Office of Cybersecurity Energy Security and Emergency Response (CESER) through the Risk Management Tools and Technologies (RMT) Program, and in part by the Laboratory Directed Research and Development Program at Oak Ridge National Laboratory (ORNL) under US DOE Grant No. DE-FG2-13ER41967. We also thank P. G. Evans for discussions and feedback. AG and GS 
acknowledge support from the Army Research Office under award W911NF-19-1-0397, and NSF under grant DGE-2152168. ORNL is managed by UT-Battelle, LLC, for the US DOE under Contract No. DE-AC05-00OR22725. The US Government retains and the publisher, by accepting the article for publication, acknowledges that the US Government retains a nonexclusive, paid-up, irrevocable, worldwide license to publish or reproduce the published form of this manuscript or allow others to do so, for the US Government purposes.

\begin{thebibliography}{58}%
	\makeatletter
	\providecommand \@ifxundefined [1]{%
		\@ifx{#1\undefined}
	}%
	\providecommand \@ifnum [1]{%
		\ifnum #1\expandafter \@firstoftwo
		\else \expandafter \@secondoftwo
		\fi
	}%
	\providecommand \@ifx [1]{%
		\ifx #1\expandafter \@firstoftwo
		\else \expandafter \@secondoftwo
		\fi
	}%
	\providecommand \natexlab [1]{#1}%
	\providecommand \enquote  [1]{``#1''}%
	\providecommand \bibnamefont  [1]{#1}%
	\providecommand \bibfnamefont [1]{#1}%
	\providecommand \citenamefont [1]{#1}%
	\providecommand \href@noop [0]{\@secondoftwo}%
	\providecommand \href [0]{\begingroup \@sanitize@url \@href}%
	\providecommand \@href[1]{\@@startlink{#1}\@@href}%
	\providecommand \@@href[1]{\endgroup#1\@@endlink}%
	\providecommand \@sanitize@url [0]{\catcode `\\12\catcode `\$12\catcode
		`\&12\catcode `\#12\catcode `\^12\catcode `\_12\catcode `\%12\relax}%
	\providecommand \@@startlink[1]{}%
	\providecommand \@@endlink[0]{}%
	\providecommand \url  [0]{\begingroup\@sanitize@url \@url }%
	\providecommand \@url [1]{\endgroup\@href {#1}{\urlprefix }}%
	\providecommand \urlprefix  [0]{URL }%
	\providecommand \Eprint [0]{\href }%
	\providecommand \doibase [0]{http://dx.doi.org/}%
	\providecommand \selectlanguage [0]{\@gobble}%
	\providecommand \bibinfo  [0]{\@secondoftwo}%
	\providecommand \bibfield  [0]{\@secondoftwo}%
	\providecommand \translation [1]{[#1]}%
	\providecommand \BibitemOpen [0]{}%
	\providecommand \bibitemStop [0]{}%
	\providecommand \bibitemNoStop [0]{.\EOS\space}%
	\providecommand \EOS [0]{\spacefactor3000\relax}%
	\providecommand \BibitemShut  [1]{\csname bibitem#1\endcsname}%
	\let\auto@bib@innerbib\@empty
	\bibitem [{\citenamefont {Rass}\ \emph {et~al.}(2020)\citenamefont {Rass},
		\citenamefont {Schauer}, \citenamefont {K{\"o}nig},\ and\ \citenamefont
		{Zhu}}]{rass2020cyber}%
	\BibitemOpen
	\bibfield  {author} {\bibinfo {author} {\bibfnamefont {S.}~\bibnamefont
			{Rass}}, \bibinfo {author} {\bibfnamefont {S.}~\bibnamefont {Schauer}},
		\bibinfo {author} {\bibfnamefont {S.}~\bibnamefont {K{\"o}nig}}, \ and\
		\bibinfo {author} {\bibfnamefont {Q.}~\bibnamefont {Zhu}},\ }\href@noop {}
	{\emph {\bibinfo {title} {Cyber-security in critical infrastructures}}},\
	Vol.\ \bibinfo {volume} {297}\ (\bibinfo  {publisher} {Springer},\ \bibinfo
	{year} {2020})\BibitemShut {NoStop}%
	\bibitem [{\citenamefont {Whyatt}\ \emph {et~al.}(2021)\citenamefont {Whyatt}
		\emph {et~al.}}]{whyatt2021}%
	\BibitemOpen
	\bibfield  {author} {\bibinfo {author} {\bibfnamefont {M.}~\bibnamefont
			{Whyatt}} \emph {et~al.},\ }\href@noop {} {\emph {\bibinfo {title} {Toward a
				Resilient Cybersecure Hydropower Fleet: Cybersecurity Landscape and Roadmap
				2021}}},\ \bibinfo {type} {Tech. Rep.}\ \bibinfo {number} {PNNL-32053}\
	(\bibinfo  {institution} {PNNL},\ \bibinfo {year} {2021})\BibitemShut
	{NoStop}%
	\bibitem [{cis(2019)}]{cisa2019}%
	\BibitemOpen
	\href@noop {} {\emph {\bibinfo {title} {Dams Sector Landscape}}},\ \bibinfo
	{type} {Tech. Rep.}\ (\bibinfo  {institution} {CISA, U.S. Department of
		Homeland Security, Cybersecurity and Infrastructure Security Agency},\
	\bibinfo {year} {2019})\BibitemShut {NoStop}%
	\bibitem [{\citenamefont {Singh}\ \emph {et~al.}(2020)\citenamefont {Singh}
		\emph {et~al.}}]{singh2020}%
	\BibitemOpen
	\bibfield  {author} {\bibinfo {author} {\bibfnamefont {P.}~\bibnamefont
			{Singh}} \emph {et~al.},\ }\href@noop {} {\bibfield  {journal} {\bibinfo
			{journal} {Materials Today: Proceedings}\ }\textbf {\bibinfo {volume} {28}},\
		\bibinfo {pages} {1569} (\bibinfo {year} {2020})}\BibitemShut {NoStop}%
	\bibitem [{\citenamefont {Ratnam}\ \emph {et~al.}(2020)\citenamefont {Ratnam},
		\citenamefont {Baldwin}, \citenamefont {Mancarella}, \citenamefont {Howden},\
		and\ \citenamefont {Seebeck}}]{RATNAM2020106833}%
	\BibitemOpen
	\bibfield  {author} {\bibinfo {author} {\bibfnamefont {E.~L.}\ \bibnamefont
			{Ratnam}}, \bibinfo {author} {\bibfnamefont {K.~G.}\ \bibnamefont {Baldwin}},
		\bibinfo {author} {\bibfnamefont {P.}~\bibnamefont {Mancarella}}, \bibinfo
		{author} {\bibfnamefont {M.}~\bibnamefont {Howden}}, \ and\ \bibinfo {author}
		{\bibfnamefont {L.}~\bibnamefont {Seebeck}},\ }\href {\doibase
		https://doi.org/10.1016/j.tej.2020.106833} {\bibfield  {journal} {\bibinfo
			{journal} {The Electricity Journal}\ }\textbf {\bibinfo {volume} {33}},\
		\bibinfo {pages} {106833} (\bibinfo {year} {2020})},\ \bibinfo {note}
	{special Issue: The Future Electricity Market Summit}\BibitemShut {NoStop}%
	\bibitem [{\citenamefont {Chen}\ and\ \citenamefont
		{Abu-Nimeh}(2011)}]{chen2011lessons}%
	\BibitemOpen
	\bibfield  {author} {\bibinfo {author} {\bibfnamefont {T.~M.}\ \bibnamefont
			{Chen}}\ and\ \bibinfo {author} {\bibfnamefont {S.}~\bibnamefont
			{Abu-Nimeh}},\ }\href@noop {} {\bibfield  {journal} {\bibinfo  {journal}
			{Computer}\ }\textbf {\bibinfo {volume} {44}},\ \bibinfo {pages} {91}
		(\bibinfo {year} {2011})}\BibitemShut {NoStop}%
	\bibitem [{\citenamefont {Passian}\ and\ \citenamefont
		{Imam}(2019)}]{passian2019}%
	\BibitemOpen
	\bibfield  {author} {\bibinfo {author} {\bibfnamefont {A.}~\bibnamefont
			{Passian}}\ and\ \bibinfo {author} {\bibfnamefont {N.}~\bibnamefont {Imam}},\
	}\href@noop {} {\bibfield  {journal} {\bibinfo  {journal} {Sensors (Basel,
				Switzerland)}\ }\textbf {\bibinfo {volume} {19}},\ \bibinfo {pages} {4048}
		(\bibinfo {year} {2019})}\BibitemShut {NoStop}%
	\bibitem [{\citenamefont {Farahi}\ \emph {et~al.}(2012)\citenamefont {Farahi}
		\emph {et~al.}}]{farahi2012}%
	\BibitemOpen
	\bibfield  {author} {\bibinfo {author} {\bibfnamefont {R.}~\bibnamefont
			{Farahi}} \emph {et~al.},\ }\href@noop {} {\bibfield  {journal} {\bibinfo
			{journal} {ACS Nano}\ }\textbf {\bibinfo {volume} {6}},\ \bibinfo {pages}
		{4548} (\bibinfo {year} {2012})}\BibitemShut {NoStop}%
	\bibitem [{\citenamefont {Alshowkan}\ \emph {et~al.}(2022)\citenamefont
		{Alshowkan}, \citenamefont {Evans}, \citenamefont {Starke}, \citenamefont
		{Earl},\ and\ \citenamefont {Peters}}]{Alshowkan2022}%
	\BibitemOpen
	\bibfield  {author} {\bibinfo {author} {\bibfnamefont {M.}~\bibnamefont
			{Alshowkan}}, \bibinfo {author} {\bibfnamefont {P.~G.}\ \bibnamefont
			{Evans}}, \bibinfo {author} {\bibfnamefont {M.}~\bibnamefont {Starke}},
		\bibinfo {author} {\bibfnamefont {D.}~\bibnamefont {Earl}}, \ and\ \bibinfo
		{author} {\bibfnamefont {N.~A.}\ \bibnamefont {Peters}},\ }\href {\doibase
		10.1038/s41598-022-16090-w} {\bibfield  {journal} {\bibinfo  {journal}
			{Scientific Reports}\ }\textbf {\bibinfo {volume} {12}},\ \bibinfo {pages}
		{12731} (\bibinfo {year} {2022})}\BibitemShut {NoStop}%
	\bibitem [{\citenamefont {Evans}\ \emph {et~al.}(2021)\citenamefont {Evans}
		\emph {et~al.}}]{evans2021}%
	\BibitemOpen
	\bibfield  {author} {\bibinfo {author} {\bibfnamefont {P.}~\bibnamefont
			{Evans}} \emph {et~al.},\ }\href@noop {} {\bibfield  {journal} {\bibinfo
			{journal} {IEEE Access}\ }\textbf {\bibinfo {volume} {9}},\ \bibinfo {pages}
		{105220} (\bibinfo {year} {2021})}\BibitemShut {NoStop}%
	\bibitem [{\citenamefont {Grice}\ \emph {et~al.}(2013)\citenamefont {Grice},
		\citenamefont {Evans},\ and\ \citenamefont {Pooser}}]{grice2013}%
	\BibitemOpen
	\bibfield  {author} {\bibinfo {author} {\bibfnamefont {W.}~\bibnamefont
			{Grice}}, \bibinfo {author} {\bibfnamefont {P.}~\bibnamefont {Evans}}, \ and\
		\bibinfo {author} {\bibfnamefont {R.}~\bibnamefont {Pooser}},\ }\enquote
	{\bibinfo {title} {Quantum key distribution for the smart grid},}\ in\
	\href@noop {} {\emph {\bibinfo {booktitle} {IEEE Vision for Smart Grid
				Communications: 2030 and Beyond}}}\ (\bibinfo {year} {2013})\BibitemShut
	{NoStop}%
	\bibitem [{\citenamefont {Kuruganti}\ \emph {et~al.}(2014)\citenamefont
		{Kuruganti} \emph {et~al.}}]{kuruganti2014}%
	\BibitemOpen
	\bibfield  {author} {\bibinfo {author} {\bibfnamefont {T.}~\bibnamefont
			{Kuruganti}} \emph {et~al.},\ }\href@noop {} {\emph {\bibinfo {title}
			{Quantum Key Distribution Applicability to Smart Grid Cybersecurity
				Systems}}},\ \bibinfo {type} {Tech. Rep.}\ (\bibinfo  {institution} {ORNL},\
	\bibinfo {year} {2014})\ \bibinfo {note} {gridSQuARe Project}\BibitemShut
	{NoStop}%
	\bibitem [{\citenamefont {Scarani}\ \emph
		{et~al.}(2009{\natexlab{a}})\citenamefont {Scarani}, \citenamefont
		{Bechmann-Pasquinucci}, \citenamefont {Cerf}, \citenamefont {Du{\v{s}}ek},
		\citenamefont {L{\"u}tkenhaus},\ and\ \citenamefont
		{Peev}}]{scarani2009security}%
	\BibitemOpen
	\bibfield  {author} {\bibinfo {author} {\bibfnamefont {V.}~\bibnamefont
			{Scarani}}, \bibinfo {author} {\bibfnamefont {H.}~\bibnamefont
			{Bechmann-Pasquinucci}}, \bibinfo {author} {\bibfnamefont {N.~J.}\
			\bibnamefont {Cerf}}, \bibinfo {author} {\bibfnamefont {M.}~\bibnamefont
			{Du{\v{s}}ek}}, \bibinfo {author} {\bibfnamefont {N.}~\bibnamefont
			{L{\"u}tkenhaus}}, \ and\ \bibinfo {author} {\bibfnamefont {M.}~\bibnamefont
			{Peev}},\ }\href@noop {} {\bibfield  {journal} {\bibinfo  {journal} {Reviews
				of Modern Physics}\ }\textbf {\bibinfo {volume} {81}},\ \bibinfo {pages}
		{1301} (\bibinfo {year} {2009}{\natexlab{a}})}\BibitemShut {NoStop}%
	\bibitem [{\citenamefont {Gobby}\ \emph {et~al.}(2004)\citenamefont {Gobby},
		\citenamefont {Yuan},\ and\ \citenamefont {Shields}}]{gobby2004quantum}%
	\BibitemOpen
	\bibfield  {author} {\bibinfo {author} {\bibfnamefont {C.}~\bibnamefont
			{Gobby}}, \bibinfo {author} {\bibfnamefont {Z.}~\bibnamefont {Yuan}}, \ and\
		\bibinfo {author} {\bibfnamefont {A.}~\bibnamefont {Shields}},\ }\href@noop
	{} {\bibfield  {journal} {\bibinfo  {journal} {Applied physics letters}\
		}\textbf {\bibinfo {volume} {84}},\ \bibinfo {pages} {3762} (\bibinfo {year}
		{2004})}\BibitemShut {NoStop}%
	\bibitem [{\citenamefont {Yuan}\ \emph {et~al.}(2007)\citenamefont {Yuan},
		\citenamefont {Kardynal}, \citenamefont {Sharpe},\ and\ \citenamefont
		{Shields}}]{yuan2007high}%
	\BibitemOpen
	\bibfield  {author} {\bibinfo {author} {\bibfnamefont {Z.}~\bibnamefont
			{Yuan}}, \bibinfo {author} {\bibfnamefont {B.}~\bibnamefont {Kardynal}},
		\bibinfo {author} {\bibfnamefont {A.}~\bibnamefont {Sharpe}}, \ and\ \bibinfo
		{author} {\bibfnamefont {A.}~\bibnamefont {Shields}},\ }\href@noop {}
	{\bibfield  {journal} {\bibinfo  {journal} {Applied Physics Letters}\
		}\textbf {\bibinfo {volume} {91}},\ \bibinfo {pages} {041114} (\bibinfo
		{year} {2007})}\BibitemShut {NoStop}%
	\bibitem [{\citenamefont {Rosenberg}\ \emph {et~al.}(2005)\citenamefont
		{Rosenberg}, \citenamefont {Peterson}, \citenamefont {Harrington},
		\citenamefont {Rice}, \citenamefont {Nakassis}, \citenamefont {Dallmann}
		\emph {et~al.}}]{rosenberg2005practical}%
	\BibitemOpen
	\bibfield  {author} {\bibinfo {author} {\bibfnamefont {D.}~\bibnamefont
			{Rosenberg}}, \bibinfo {author} {\bibfnamefont {C.}~\bibnamefont {Peterson}},
		\bibinfo {author} {\bibfnamefont {J.}~\bibnamefont {Harrington}}, \bibinfo
		{author} {\bibfnamefont {P.}~\bibnamefont {Rice}}, \bibinfo {author}
		{\bibfnamefont {A.}~\bibnamefont {Nakassis}}, \bibinfo {author}
		{\bibfnamefont {N.}~\bibnamefont {Dallmann}},  \emph {et~al.},\ }\href@noop
	{} {\bibfield  {journal} {\bibinfo  {journal} {New Journal of Physics}\
		}\textbf {\bibinfo {volume} {7}},\ \bibinfo {pages} {71} (\bibinfo {year}
		{2005})}\BibitemShut {NoStop}%
	\bibitem [{\citenamefont {Hiskett}\ \emph {et~al.}(2007)\citenamefont
		{Hiskett}, \citenamefont {Buller}, \citenamefont {Loudon}, \citenamefont
		{Matthews}, \citenamefont {Pryde}, \citenamefont {Gilchrist} \emph
		{et~al.}}]{hiskett2007long}%
	\BibitemOpen
	\bibfield  {author} {\bibinfo {author} {\bibfnamefont {P.}~\bibnamefont
			{Hiskett}}, \bibinfo {author} {\bibfnamefont {G.}~\bibnamefont {Buller}},
		\bibinfo {author} {\bibfnamefont {A.}~\bibnamefont {Loudon}}, \bibinfo
		{author} {\bibfnamefont {J.}~\bibnamefont {Matthews}}, \bibinfo {author}
		{\bibfnamefont {G.}~\bibnamefont {Pryde}}, \bibinfo {author} {\bibfnamefont
			{A.}~\bibnamefont {Gilchrist}},  \emph {et~al.},\ }\href@noop {} {\bibfield
		{journal} {\bibinfo  {journal} {New Journal of Physics}\ }\textbf {\bibinfo
			{volume} {8}},\ \bibinfo {pages} {193} (\bibinfo {year} {2007})}\BibitemShut
	{NoStop}%
	\bibitem [{\citenamefont {Lawrence}\ \emph {et~al.}(2020)\citenamefont
		{Lawrence} \emph {et~al.}}]{lawrence2020}%
	\BibitemOpen
	\bibfield  {author} {\bibinfo {author} {\bibfnamefont {J.}~\bibnamefont
			{Lawrence}} \emph {et~al.},\ }\href {\doibase 10.2172/1764035} {\emph
		{\bibinfo {title} {Fossil Power Plant Cyber Security Life-Cycle Risk
				Reduction, A Practical Framework for Implementation}}},\ \bibinfo {type}
	{Tech. Rep.}\ (\bibinfo  {institution} {United States},\ \bibinfo {year}
	{2020})\BibitemShut {NoStop}%
	\bibitem [{\citenamefont {Pakki Bharani~Chandra}\ and\ \citenamefont
		{Potnuru}(2021)}]{chandra2021}%
	\BibitemOpen
	\bibfield  {author} {\bibinfo {author} {\bibfnamefont {K.}~\bibnamefont
			{Pakki Bharani~Chandra}}\ and\ \bibinfo {author} {\bibfnamefont
			{D.}~\bibnamefont {Potnuru}},\ }\href@noop {} {\bibfield  {journal} {\bibinfo
			{journal} {International Journal of Ambient Energy}\ }\textbf {\bibinfo
			{volume} {42}},\ \bibinfo {pages} {203} (\bibinfo {year} {2021})}\BibitemShut
	{NoStop}%
	\bibitem [{\citenamefont {Wang}\ \emph {et~al.}(2022)\citenamefont {Wang} \emph
		{et~al.}}]{wang2022}%
	\BibitemOpen
	\bibfield  {author} {\bibinfo {author} {\bibfnamefont {S.}~\bibnamefont
			{Wang}} \emph {et~al.},\ }\href@noop {} {\bibfield  {journal} {\bibinfo
			{journal} {Nature Photonics}\ }\textbf {\bibinfo {volume} {16}},\ \bibinfo
		{pages} {154} (\bibinfo {year} {2022})}\BibitemShut {NoStop}%
	\bibitem [{\citenamefont {Liao}\ \emph {et~al.}(2017)\citenamefont {Liao} \emph
		{et~al.}}]{liao2017}%
	\BibitemOpen
	\bibfield  {author} {\bibinfo {author} {\bibfnamefont {S.}~\bibnamefont
			{Liao}} \emph {et~al.},\ }\href@noop {} {\bibfield  {journal} {\bibinfo
			{journal} {Nature}\ }\textbf {\bibinfo {volume} {549}},\ \bibinfo {pages}
		{43} (\bibinfo {year} {2017})}\BibitemShut {NoStop}%
	\bibitem [{\citenamefont {Yin}\ \emph {et~al.}(2020)\citenamefont {Yin} \emph
		{et~al.}}]{yin2020}%
	\BibitemOpen
	\bibfield  {author} {\bibinfo {author} {\bibfnamefont {J.}~\bibnamefont
			{Yin}} \emph {et~al.},\ }\href@noop {} {\bibfield  {journal} {\bibinfo
			{journal} {Nature}\ }\textbf {\bibinfo {volume} {582}},\ \bibinfo {pages}
		{501} (\bibinfo {year} {2020})}\BibitemShut {NoStop}%
	\bibitem [{\citenamefont {Sharma}\ \emph {et~al.}(2021)\citenamefont {Sharma},
		\citenamefont {Agrawal}, \citenamefont {Bhatia}, \citenamefont {Prakash},\
		and\ \citenamefont {Mishra}}]{sharma9520678}%
	\BibitemOpen
	\bibfield  {author} {\bibinfo {author} {\bibfnamefont {P.}~\bibnamefont
			{Sharma}}, \bibinfo {author} {\bibfnamefont {A.}~\bibnamefont {Agrawal}},
		\bibinfo {author} {\bibfnamefont {V.}~\bibnamefont {Bhatia}}, \bibinfo
		{author} {\bibfnamefont {S.}~\bibnamefont {Prakash}}, \ and\ \bibinfo
		{author} {\bibfnamefont {A.~K.}\ \bibnamefont {Mishra}},\ }\href {\doibase
		10.1109/OJCOMS.2021.3106659} {\bibfield  {journal} {\bibinfo  {journal} {IEEE
				Open Journal of the Communications Society}\ }\textbf {\bibinfo {volume}
			{2}},\ \bibinfo {pages} {2049} (\bibinfo {year} {2021})}\BibitemShut
	{NoStop}%
	\bibitem [{\citenamefont {Scarani}\ \emph
		{et~al.}(2009{\natexlab{b}})\citenamefont {Scarani} \emph
		{et~al.}}]{scarani2009}%
	\BibitemOpen
	\bibfield  {author} {\bibinfo {author} {\bibfnamefont {V.}~\bibnamefont
			{Scarani}} \emph {et~al.},\ }\href@noop {} {\bibfield  {journal} {\bibinfo
			{journal} {Reviews of Modern Physics}\ }\textbf {\bibinfo {volume} {81}}
		(\bibinfo {year} {2009}{\natexlab{b}})}\BibitemShut {NoStop}%
	\bibitem [{\citenamefont {Qi}\ \emph {et~al.}(2010)\citenamefont {Qi},
		\citenamefont {Zhu}, \citenamefont {Qian},\ and\ \citenamefont
		{Lo}}]{qi2010}%
	\BibitemOpen
	\bibfield  {author} {\bibinfo {author} {\bibfnamefont {B.}~\bibnamefont
			{Qi}}, \bibinfo {author} {\bibfnamefont {W.}~\bibnamefont {Zhu}}, \bibinfo
		{author} {\bibfnamefont {L.}~\bibnamefont {Qian}}, \ and\ \bibinfo {author}
		{\bibfnamefont {H.-K.}\ \bibnamefont {Lo}},\ }\href {\doibase
		10.1088/1367-2630/12/10/103042} {\bibfield  {journal} {\bibinfo  {journal}
			{New Journal of Physics}\ }\textbf {\bibinfo {volume} {12}},\ \bibinfo
		{pages} {103042} (\bibinfo {year} {2010})}\BibitemShut {NoStop}%
	\bibitem [{\citenamefont {Scherer}\ \emph {et~al.}(2011)\citenamefont {Scherer}
		\emph {et~al.}}]{Scherer2011}%
	\BibitemOpen
	\bibfield  {author} {\bibinfo {author} {\bibfnamefont {A.}~\bibnamefont
			{Scherer}} \emph {et~al.},\ }\href@noop {} {\bibfield  {journal} {\bibinfo
			{journal} {Optics express}\ }\textbf {\bibinfo {volume} {19}} (\bibinfo
		{year} {2011})}\BibitemShut {NoStop}%
	\bibitem [{\citenamefont {Lo}\ \emph {et~al.}(2005)\citenamefont {Lo} \emph
		{et~al.}}]{Lo2005}%
	\BibitemOpen
	\bibfield  {author} {\bibinfo {author} {\bibfnamefont {H.~K.}\ \bibnamefont
			{Lo}} \emph {et~al.},\ }\href@noop {} {\bibfield  {journal} {\bibinfo
			{journal} {Physical Review Letters}\ }\textbf {\bibinfo {volume} {94}}
		(\bibinfo {year} {2005})}\BibitemShut {NoStop}%
	\bibitem [{\citenamefont {Wolf}(2021)}]{ramona:book}%
	\BibitemOpen
	\bibfield  {author} {\bibinfo {author} {\bibfnamefont {R.}~\bibnamefont
			{Wolf}},\ }\href@noop {} {\emph {\bibinfo {title} {Quantum key
				distribution}}}\ (\bibinfo  {publisher} {Springer},\ \bibinfo {year}
	{2021})\BibitemShut {NoStop}%
	\bibitem [{\citenamefont {L{\"u}tkenhaus}(2000)}]{lutkenhaus2000security}%
	\BibitemOpen
	\bibfield  {author} {\bibinfo {author} {\bibfnamefont {N.}~\bibnamefont
			{L{\"u}tkenhaus}},\ }\href@noop {} {\bibfield  {journal} {\bibinfo  {journal}
			{Physical Review A}\ }\textbf {\bibinfo {volume} {61}},\ \bibinfo {pages}
		{052304} (\bibinfo {year} {2000})}\BibitemShut {NoStop}%
	\bibitem [{\citenamefont {Makarov}\ and\ \citenamefont
		{Khan}(2006)}]{makarov2006optical}%
	\BibitemOpen
	\bibfield  {author} {\bibinfo {author} {\bibfnamefont {V.}~\bibnamefont
			{Makarov}}\ and\ \bibinfo {author} {\bibfnamefont {J.}~\bibnamefont {Khan}},\
	}\href@noop {} {\bibfield  {journal} {\bibinfo  {journal} {Optics letters}\
		}\textbf {\bibinfo {volume} {30}},\ \bibinfo {pages} {1043} (\bibinfo {year}
		{2006})}\BibitemShut {NoStop}%
	\bibitem [{\citenamefont {Pljonkin}\ and\ \citenamefont
		{Singh}(2018)}]{Anton_review}%
	\BibitemOpen
	\bibfield  {author} {\bibinfo {author} {\bibfnamefont {A.}~\bibnamefont
			{Pljonkin}}\ and\ \bibinfo {author} {\bibfnamefont {P.~K.}\ \bibnamefont
			{Singh}},\ }in\ \href {\doibase 10.1109/PDGC.2018.8745822} {\emph {\bibinfo
			{booktitle} {2018 Fifth International Conference on Parallel, Distributed and
				Grid Computing (PDGC)}}}\ (\bibinfo {year} {2018})\ pp.\ \bibinfo {pages}
	{795--799}\BibitemShut {NoStop}%
	\bibitem [{\citenamefont {Moreno~Escobar}\ \emph {et~al.}(2021)\citenamefont
		{Moreno~Escobar}, \citenamefont {Morales~Matamoros}, \citenamefont
		{Tejeida~Padilla}, \citenamefont {Lina~Reyes},\ and\ \citenamefont
		{Quintana~Espinosa}}]{moreno2021comprehensive}%
	\BibitemOpen
	\bibfield  {author} {\bibinfo {author} {\bibfnamefont {J.~J.}\ \bibnamefont
			{Moreno~Escobar}}, \bibinfo {author} {\bibfnamefont {O.}~\bibnamefont
			{Morales~Matamoros}}, \bibinfo {author} {\bibfnamefont {R.}~\bibnamefont
			{Tejeida~Padilla}}, \bibinfo {author} {\bibfnamefont {I.}~\bibnamefont
			{Lina~Reyes}}, \ and\ \bibinfo {author} {\bibfnamefont {H.}~\bibnamefont
			{Quintana~Espinosa}},\ }\href@noop {} {\bibfield  {journal} {\bibinfo
			{journal} {Sensors}\ }\textbf {\bibinfo {volume} {21}},\ \bibinfo {pages}
		{6978} (\bibinfo {year} {2021})}\BibitemShut {NoStop}%
	\bibitem [{\citenamefont {Kong}(2020)}]{kong2020review}%
	\BibitemOpen
	\bibfield  {author} {\bibinfo {author} {\bibfnamefont {P.-Y.}\ \bibnamefont
			{Kong}},\ }\href@noop {} {\bibfield  {journal} {\bibinfo  {journal} {IEEE
				Systems Journal}\ }\textbf {\bibinfo {volume} {16}},\ \bibinfo {pages} {41}
		(\bibinfo {year} {2020})}\BibitemShut {NoStop}%
	\bibitem [{\citenamefont {Gopstein}\ \emph {et~al.}(2021)\citenamefont
		{Gopstein} \emph {et~al.}}]{gopstein2021}%
	\BibitemOpen
	\bibfield  {author} {\bibinfo {author} {\bibfnamefont {A.}~\bibnamefont
			{Gopstein}} \emph {et~al.},\ }\href
	{https://tsapps.nist.gov/publication/get_pdf.cfm?pub_id=931882} {\enquote
		{\bibinfo {title} {Nist framework and roadmap for smart grid interoperability
				standards, release 2 (latest 4)},}\ } (\bibinfo {year} {2021})\BibitemShut
	{NoStop}%
	\bibitem [{ces(2011)}]{ceser2011}%
	\BibitemOpen
	\href
	{https://www.energy.gov/ceser/downloads/21-steps-improve-cyber-security-scada-networks}
	{\enquote {\bibinfo {title} {21 steps to improve cyber security of scada
				network},}\ } (\bibinfo {year} {2011})\BibitemShut {NoStop}%
	\bibitem [{\citenamefont {Alrefaei}(2022)}]{Alrefaei2022AnOO}%
	\BibitemOpen
	\bibfield  {author} {\bibinfo {author} {\bibfnamefont {A.~S.}\ \bibnamefont
			{Alrefaei}},\ }\href {https://api.semanticscholar.org/CorpusID:257652020}
	{\bibfield  {journal} {\bibinfo  {journal} {2022 Fifth National Conference of
				Saudi Computers Colleges (NCCC)}\ ,\ \bibinfo {pages} {88}} (\bibinfo {year}
		{2022})}\BibitemShut {NoStop}%
	\bibitem [{\citenamefont {Lim}\ \emph {et~al.}(2014)\citenamefont {Lim} \emph
		{et~al.}}]{lim2014}%
	\BibitemOpen
	\bibfield  {author} {\bibinfo {author} {\bibfnamefont {C.}~\bibnamefont
			{Lim}} \emph {et~al.},\ }\href@noop {} {\bibfield  {journal} {\bibinfo
			{journal} {Physical Review A}\ }\textbf {\bibinfo {volume} {89}} (\bibinfo
		{year} {2014})}\BibitemShut {NoStop}%
	\bibitem [{\citenamefont {Wilde}(2017)}]{wilde_2017}%
	\BibitemOpen
	\bibfield  {author} {\bibinfo {author} {\bibfnamefont {M.~M.}\ \bibnamefont
			{Wilde}},\ }\enquote {\bibinfo {title} {Preface to the second edition},}\ in\
	\href {\doibase 10.1017/9781316809976.001} {\emph {\bibinfo {booktitle}
			{Quantum Information Theory}}}\ (\bibinfo  {publisher} {Cambridge University
		Press},\ \bibinfo {year} {2017})\ pp.\ \bibinfo {pages} {xi--xii},\ \bibinfo
	{edition} {2nd}\ ed.\BibitemShut {Stop}%
	\bibitem [{\citenamefont {Moschandreou}\ \emph {et~al.}(2021)\citenamefont
		{Moschandreou} \emph {et~al.}}]{Moschandreou2021}%
	\BibitemOpen
	\bibfield  {author} {\bibinfo {author} {\bibfnamefont {E.}~\bibnamefont
			{Moschandreou}} \emph {et~al.},\ }\href@noop {} {\bibfield  {journal}
		{\bibinfo  {journal} {Physical Review A}\ }\textbf {\bibinfo {volume} {103}}
		(\bibinfo {year} {2021})}\BibitemShut {NoStop}%
	\bibitem [{\citenamefont {Ding}\ \emph {et~al.}(2017)\citenamefont {Ding},
		\citenamefont {Chen}, \citenamefont {Wang}, \citenamefont {He}, \citenamefont
		{Yin}, \citenamefont {Chen}, \citenamefont {Zhou}, \citenamefont {Guo},\ and\
		\citenamefont {Han}}]{ding2017polarization}%
	\BibitemOpen
	\bibfield  {author} {\bibinfo {author} {\bibfnamefont {Y.-Y.}\ \bibnamefont
			{Ding}}, \bibinfo {author} {\bibfnamefont {H.}~\bibnamefont {Chen}}, \bibinfo
		{author} {\bibfnamefont {S.}~\bibnamefont {Wang}}, \bibinfo {author}
		{\bibfnamefont {D.-Y.}\ \bibnamefont {He}}, \bibinfo {author} {\bibfnamefont
			{Z.-Q.}\ \bibnamefont {Yin}}, \bibinfo {author} {\bibfnamefont
			{W.}~\bibnamefont {Chen}}, \bibinfo {author} {\bibfnamefont {Z.}~\bibnamefont
			{Zhou}}, \bibinfo {author} {\bibfnamefont {G.-C.}\ \bibnamefont {Guo}}, \
		and\ \bibinfo {author} {\bibfnamefont {Z.-F.}\ \bibnamefont {Han}},\
	}\href@noop {} {\bibfield  {journal} {\bibinfo  {journal} {Optics Express}\
		}\textbf {\bibinfo {volume} {25}},\ \bibinfo {pages} {27923} (\bibinfo {year}
		{2017})}\BibitemShut {NoStop}%
	\bibitem [{\citenamefont {Mekhtiev}\ \emph {et~al.}(2021)\citenamefont
		{Mekhtiev}, \citenamefont {Gerasin}, \citenamefont {Rudavin}, \citenamefont
		{Duplinsky},\ and\ \citenamefont {Kurochkin}}]{mekhtiev2021polarization}%
	\BibitemOpen
	\bibfield  {author} {\bibinfo {author} {\bibfnamefont {E.}~\bibnamefont
			{Mekhtiev}}, \bibinfo {author} {\bibfnamefont {I.}~\bibnamefont {Gerasin}},
		\bibinfo {author} {\bibfnamefont {N.}~\bibnamefont {Rudavin}}, \bibinfo
		{author} {\bibfnamefont {A.}~\bibnamefont {Duplinsky}}, \ and\ \bibinfo
		{author} {\bibfnamefont {Y.}~\bibnamefont {Kurochkin}},\ }in\ \href@noop {}
	{\emph {\bibinfo {booktitle} {Journal of Physics: Conference Series}}},\
	Vol.\ \bibinfo {volume} {2086}\ (\bibinfo {organization} {IOP Publishing},\
	\bibinfo {year} {2021})\ p.\ \bibinfo {pages} {012092}\BibitemShut {NoStop}%
	\bibitem [{\citenamefont {Wu}\ \emph {et~al.}(2006)\citenamefont {Wu},
		\citenamefont {Chen}, \citenamefont {Li},\ and\ \citenamefont
		{Zeng}}]{wu2006stable}%
	\BibitemOpen
	\bibfield  {author} {\bibinfo {author} {\bibfnamefont {G.}~\bibnamefont
			{Wu}}, \bibinfo {author} {\bibfnamefont {J.}~\bibnamefont {Chen}}, \bibinfo
		{author} {\bibfnamefont {Y.}~\bibnamefont {Li}}, \ and\ \bibinfo {author}
		{\bibfnamefont {H.}~\bibnamefont {Zeng}},\ }\href@noop {} {\bibfield
		{journal} {\bibinfo  {journal} {arXiv preprint quant-ph/0606108}\ } (\bibinfo
		{year} {2006})}\BibitemShut {NoStop}%
	\bibitem [{\citenamefont {Inaudi}\ and\ \citenamefont
		{Blin}()}]{inaudimonitoring}%
	\BibitemOpen
	\bibfield  {author} {\bibinfo {author} {\bibfnamefont {D.}~\bibnamefont
			{Inaudi}}\ and\ \bibinfo {author} {\bibfnamefont {E.~R.}\ \bibnamefont
			{Blin}},\ }\href@noop {} {\ }\BibitemShut {NoStop}%
	\bibitem [{\citenamefont {de~la Torre}\ \emph {et~al.}(2021)\citenamefont
		{de~la Torre}, \citenamefont {Floris}, \citenamefont {Sales},\ and\
		\citenamefont {Escaler}}]{de2021fiber}%
	\BibitemOpen
	\bibfield  {author} {\bibinfo {author} {\bibfnamefont {O.}~\bibnamefont
			{de~la Torre}}, \bibinfo {author} {\bibfnamefont {I.}~\bibnamefont {Floris}},
		\bibinfo {author} {\bibfnamefont {S.}~\bibnamefont {Sales}}, \ and\ \bibinfo
		{author} {\bibfnamefont {X.}~\bibnamefont {Escaler}},\ }\href@noop {}
	{\bibfield  {journal} {\bibinfo  {journal} {Sensors}\ }\textbf {\bibinfo
			{volume} {21}},\ \bibinfo {pages} {4272} (\bibinfo {year}
		{2021})}\BibitemShut {NoStop}%
	\bibitem [{\citenamefont {Peters}\ \emph {et~al.}(2009)\citenamefont {Peters},
		\citenamefont {Toliver}, \citenamefont {Chapuran}, \citenamefont {Runser},
		\citenamefont {McNown}, \citenamefont {Peterson}, \citenamefont {Rosenberg},
		\citenamefont {Dallmann}, \citenamefont {Hughes}, \citenamefont {McCabe},
		\citenamefont {Nordholt},\ and\ \citenamefont {Tyagi}}]{Peters_2009}%
	\BibitemOpen
	\bibfield  {author} {\bibinfo {author} {\bibfnamefont {N.~A.}\ \bibnamefont
			{Peters}}, \bibinfo {author} {\bibfnamefont {P.}~\bibnamefont {Toliver}},
		\bibinfo {author} {\bibfnamefont {T.~E.}\ \bibnamefont {Chapuran}}, \bibinfo
		{author} {\bibfnamefont {R.~J.}\ \bibnamefont {Runser}}, \bibinfo {author}
		{\bibfnamefont {S.~R.}\ \bibnamefont {McNown}}, \bibinfo {author}
		{\bibfnamefont {C.~G.}\ \bibnamefont {Peterson}}, \bibinfo {author}
		{\bibfnamefont {D.}~\bibnamefont {Rosenberg}}, \bibinfo {author}
		{\bibfnamefont {N.}~\bibnamefont {Dallmann}}, \bibinfo {author}
		{\bibfnamefont {R.~J.}\ \bibnamefont {Hughes}}, \bibinfo {author}
		{\bibfnamefont {K.~P.}\ \bibnamefont {McCabe}}, \bibinfo {author}
		{\bibfnamefont {J.~E.}\ \bibnamefont {Nordholt}}, \ and\ \bibinfo {author}
		{\bibfnamefont {K.~T.}\ \bibnamefont {Tyagi}},\ }\href {\doibase
		10.1088/1367-2630/11/4/045012} {\bibfield  {journal} {\bibinfo  {journal}
			{New Journal of Physics}\ }\textbf {\bibinfo {volume} {11}},\ \bibinfo
		{pages} {045012} (\bibinfo {year} {2009})}\BibitemShut {NoStop}%
	\bibitem [{\citenamefont {Chapuran}\ \emph {et~al.}(2009)\citenamefont
		{Chapuran}, \citenamefont {Toliver}, \citenamefont {Peters}, \citenamefont
		{Jackel}, \citenamefont {Goodman}, \citenamefont {Runser}, \citenamefont
		{McNown}, \citenamefont {Dallmann}, \citenamefont {Hughes}, \citenamefont
		{McCabe}, \citenamefont {Nordholt}, \citenamefont {Peterson}, \citenamefont
		{Tyagi}, \citenamefont {Mercer},\ and\ \citenamefont
		{Dardy}}]{Chapuran_2009}%
	\BibitemOpen
	\bibfield  {author} {\bibinfo {author} {\bibfnamefont {T.~E.}\ \bibnamefont
			{Chapuran}}, \bibinfo {author} {\bibfnamefont {P.}~\bibnamefont {Toliver}},
		\bibinfo {author} {\bibfnamefont {N.~A.}\ \bibnamefont {Peters}}, \bibinfo
		{author} {\bibfnamefont {J.}~\bibnamefont {Jackel}}, \bibinfo {author}
		{\bibfnamefont {M.~S.}\ \bibnamefont {Goodman}}, \bibinfo {author}
		{\bibfnamefont {R.~J.}\ \bibnamefont {Runser}}, \bibinfo {author}
		{\bibfnamefont {S.~R.}\ \bibnamefont {McNown}}, \bibinfo {author}
		{\bibfnamefont {N.}~\bibnamefont {Dallmann}}, \bibinfo {author}
		{\bibfnamefont {R.~J.}\ \bibnamefont {Hughes}}, \bibinfo {author}
		{\bibfnamefont {K.~P.}\ \bibnamefont {McCabe}}, \bibinfo {author}
		{\bibfnamefont {J.~E.}\ \bibnamefont {Nordholt}}, \bibinfo {author}
		{\bibfnamefont {C.~G.}\ \bibnamefont {Peterson}}, \bibinfo {author}
		{\bibfnamefont {K.~T.}\ \bibnamefont {Tyagi}}, \bibinfo {author}
		{\bibfnamefont {L.}~\bibnamefont {Mercer}}, \ and\ \bibinfo {author}
		{\bibfnamefont {H.}~\bibnamefont {Dardy}},\ }\href {\doibase
		10.1088/1367-2630/11/10/105001} {\bibfield  {journal} {\bibinfo  {journal}
			{New Journal of Physics}\ }\textbf {\bibinfo {volume} {11}},\ \bibinfo
		{pages} {105001} (\bibinfo {year} {2009})}\BibitemShut {NoStop}%
	\bibitem [{\citenamefont {Bloom}\ \emph {et~al.}(2022)\citenamefont {Bloom},
		\citenamefont {Fields}, \citenamefont {Maslennikov},\ and\ \citenamefont
		{Rozenman}}]{physics4010009}%
	\BibitemOpen
	\bibfield  {author} {\bibinfo {author} {\bibfnamefont {Y.}~\bibnamefont
			{Bloom}}, \bibinfo {author} {\bibfnamefont {I.}~\bibnamefont {Fields}},
		\bibinfo {author} {\bibfnamefont {A.}~\bibnamefont {Maslennikov}}, \ and\
		\bibinfo {author} {\bibfnamefont {G.~G.}\ \bibnamefont {Rozenman}},\ }\href
	{\doibase 10.3390/physics4010009} {\bibfield  {journal} {\bibinfo  {journal}
			{Physics}\ }\textbf {\bibinfo {volume} {4}},\ \bibinfo {pages} {104}
		(\bibinfo {year} {2022})}\BibitemShut {NoStop}%
	\bibitem [{\citenamefont {Siehler}(2011)}]{siehler2011hamming}%
	\BibitemOpen
	\bibfield  {author} {\bibinfo {author} {\bibfnamefont {J.~A.}\ \bibnamefont
			{Siehler}},\ }\href {http://demonstrations.wolfram.com/TheHamming74Code/}
	{\enquote {\bibinfo {title} {The hamming(7,4) code},}\ } (\bibinfo {year}
	{2011})\BibitemShut {NoStop}%
	\bibitem [{\citenamefont {Urbina}\ \emph {et~al.}(2016)\citenamefont {Urbina},
		\citenamefont {Giraldo}, \citenamefont {Cardenas},\ and\ \citenamefont
		{Tippenhauer}}]{urbina2016survey}%
	\BibitemOpen
	\bibfield  {author} {\bibinfo {author} {\bibfnamefont {D.~I.}\ \bibnamefont
			{Urbina}}, \bibinfo {author} {\bibfnamefont {J.~A.}\ \bibnamefont {Giraldo}},
		\bibinfo {author} {\bibfnamefont {A.~A.}\ \bibnamefont {Cardenas}}, \ and\
		\bibinfo {author} {\bibfnamefont {N.~O.}\ \bibnamefont {Tippenhauer}},\ }in\
	\href@noop {} {\emph {\bibinfo {booktitle} {IFIP Annual Conference on Data
				and Applications Security and Privacy}}}\ (\bibinfo {organization} {Springer,
		Cham},\ \bibinfo {year} {2016})\ pp.\ \bibinfo {pages} {65--81}\BibitemShut
	{NoStop}%
	\bibitem [{\citenamefont {Lee}\ \emph {et~al.}(2016)\citenamefont {Lee},
		\citenamefont {Assante},\ and\ \citenamefont {Conway}}]{lee2016analysis}%
	\BibitemOpen
	\bibfield  {author} {\bibinfo {author} {\bibfnamefont {R.~M.}\ \bibnamefont
			{Lee}}, \bibinfo {author} {\bibfnamefont {M.~J.}\ \bibnamefont {Assante}}, \
		and\ \bibinfo {author} {\bibfnamefont {T.}~\bibnamefont {Conway}},\
	}\href@noop {} {\emph {\bibinfo {title} {Analysis of the Cyber Attack on the
				Ukrainian Power Grid}}},\ \bibinfo {type} {Tech. Rep.}\ (\bibinfo
	{institution} {Electricity Information Sharing and Analysis Center
		(E-ISAC)},\ \bibinfo {year} {2016})\BibitemShut {NoStop}%
	\bibitem [{\citenamefont {Ouellet}\ \emph {et~al.}(2022)\citenamefont
		{Ouellet}, \citenamefont {Dettmer}, \citenamefont {Olivier}, \citenamefont
		{DeWit},\ and\ \citenamefont {Lato}}]{ouellet2022advanced}%
	\BibitemOpen
	\bibfield  {author} {\bibinfo {author} {\bibfnamefont {S.~M.}\ \bibnamefont
			{Ouellet}}, \bibinfo {author} {\bibfnamefont {J.}~\bibnamefont {Dettmer}},
		\bibinfo {author} {\bibfnamefont {G.}~\bibnamefont {Olivier}}, \bibinfo
		{author} {\bibfnamefont {T.}~\bibnamefont {DeWit}}, \ and\ \bibinfo {author}
		{\bibfnamefont {M.}~\bibnamefont {Lato}},\ }\href@noop {} {\bibfield
		{journal} {\bibinfo  {journal} {Communications Earth \& Environment}\
		}\textbf {\bibinfo {volume} {3}},\ \bibinfo {pages} {301} (\bibinfo {year}
		{2022})}\BibitemShut {NoStop}%
	\bibitem [{\citenamefont {Antonovskaya}\ \emph {et~al.}(2019)\citenamefont
		{Antonovskaya}, \citenamefont {Kapustian}, \citenamefont {Basakina},
		\citenamefont {Afonin},\ and\ \citenamefont
		{Moshkunov}}]{geosciences9040187}%
	\BibitemOpen
	\bibfield  {author} {\bibinfo {author} {\bibfnamefont {G.}~\bibnamefont
			{Antonovskaya}}, \bibinfo {author} {\bibfnamefont {N.}~\bibnamefont
			{Kapustian}}, \bibinfo {author} {\bibfnamefont {I.}~\bibnamefont {Basakina}},
		\bibinfo {author} {\bibfnamefont {N.}~\bibnamefont {Afonin}}, \ and\ \bibinfo
		{author} {\bibfnamefont {K.}~\bibnamefont {Moshkunov}},\ }\href {\doibase
		10.3390/geosciences9040187} {\bibfield  {journal} {\bibinfo  {journal}
			{Geosciences}\ }\textbf {\bibinfo {volume} {9}} (\bibinfo {year} {2019}),\
		10.3390/geosciences9040187}\BibitemShut {NoStop}%
	\bibitem [{\citenamefont {Baron}\ \emph {et~al.}(2022)\citenamefont {Baron},
		\citenamefont {Kočiško}, \citenamefont {Hlavatá},\ and\ \citenamefont
		{Franas}}]{petr}%
	\BibitemOpen
	\bibfield  {author} {\bibinfo {author} {\bibfnamefont {P.}~\bibnamefont
			{Baron}}, \bibinfo {author} {\bibfnamefont {M.}~\bibnamefont {Kočiško}},
		\bibinfo {author} {\bibfnamefont {S.}~\bibnamefont {Hlavatá}}, \ and\
		\bibinfo {author} {\bibfnamefont {E.}~\bibnamefont {Franas}},\ }\href
	{\doibase 10.1177/16878132221101023} {\bibfield  {journal} {\bibinfo
			{journal} {Advances in Mechanical Engineering}\ }\textbf {\bibinfo {volume}
			{14}},\ \bibinfo {pages} {16878132221101023} (\bibinfo {year} {2022})},\
	\Eprint {http://arxiv.org/abs/https://doi.org/10.1177/16878132221101023}
	{https://doi.org/10.1177/16878132221101023} \BibitemShut {NoStop}%
	\bibitem [{\citenamefont {Mohanta}\ \emph {et~al.}(2017)\citenamefont
		{Mohanta}, \citenamefont {Chelliah}, \citenamefont {Allamsetty},
		\citenamefont {Akula},\ and\ \citenamefont {Ghosh}}]{MOHANTA2017637}%
	\BibitemOpen
	\bibfield  {author} {\bibinfo {author} {\bibfnamefont {R.~K.}\ \bibnamefont
			{Mohanta}}, \bibinfo {author} {\bibfnamefont {T.~R.}\ \bibnamefont
			{Chelliah}}, \bibinfo {author} {\bibfnamefont {S.}~\bibnamefont
			{Allamsetty}}, \bibinfo {author} {\bibfnamefont {A.}~\bibnamefont {Akula}}, \
		and\ \bibinfo {author} {\bibfnamefont {R.}~\bibnamefont {Ghosh}},\ }\href
	{\doibase https://doi.org/10.1016/j.jestch.2016.11.004} {\bibfield  {journal}
		{\bibinfo  {journal} {Engineering Science and Technology, an International
				Journal}\ }\textbf {\bibinfo {volume} {20}},\ \bibinfo {pages} {637}
		(\bibinfo {year} {2017})}\BibitemShut {NoStop}%
	\bibitem [{\citenamefont {Quaranta}\ and\ \citenamefont
		{M{\"u}ller}(2021)}]{quaranta2021noise}%
	\BibitemOpen
	\bibfield  {author} {\bibinfo {author} {\bibfnamefont {E.}~\bibnamefont
			{Quaranta}}\ and\ \bibinfo {author} {\bibfnamefont {G.}~\bibnamefont
			{M{\"u}ller}},\ }\href@noop {} {\bibfield  {journal} {\bibinfo  {journal}
			{International Journal of Environmental Research and Public Health}\ }\textbf
		{\bibinfo {volume} {18}},\ \bibinfo {pages} {13051} (\bibinfo {year}
		{2021})}\BibitemShut {NoStop}%
	\bibitem [{\citenamefont {Diamanti}\ \emph {et~al.}(2016)\citenamefont
		{Diamanti}, \citenamefont {Lo}, \citenamefont {Qi},\ and\ \citenamefont
		{Yuan}}]{diamanti2016practical}%
	\BibitemOpen
	\bibfield  {author} {\bibinfo {author} {\bibfnamefont {E.}~\bibnamefont
			{Diamanti}}, \bibinfo {author} {\bibfnamefont {H.-K.}\ \bibnamefont {Lo}},
		\bibinfo {author} {\bibfnamefont {B.}~\bibnamefont {Qi}}, \ and\ \bibinfo
		{author} {\bibfnamefont {Z.}~\bibnamefont {Yuan}},\ }\href@noop {} {\bibfield
		{journal} {\bibinfo  {journal} {npj Quantum Information}\ }\textbf {\bibinfo
			{volume} {2}},\ \bibinfo {pages} {1} (\bibinfo {year} {2016})}\BibitemShut
	{NoStop}%
	\bibitem [{\citenamefont {Nandal}\ \emph {et~al.}(2020)\citenamefont {Nandal},
		\citenamefont {Nandal}, \citenamefont {Joshi},\ and\ \citenamefont
		{Rathee}}]{nandal2020survey}%
	\BibitemOpen
	\bibfield  {author} {\bibinfo {author} {\bibfnamefont {R.}~\bibnamefont
			{Nandal}}, \bibinfo {author} {\bibfnamefont {A.}~\bibnamefont {Nandal}},
		\bibinfo {author} {\bibfnamefont {K.}~\bibnamefont {Joshi}}, \ and\ \bibinfo
		{author} {\bibfnamefont {A.~K.}\ \bibnamefont {Rathee}},\ }\href@noop {} {\
		(\bibinfo {year} {2020})}\BibitemShut {NoStop}%
	\bibitem [{\citenamefont {Loudon}(2000)}]{loudon2000quantum}%
	\BibitemOpen
	\bibfield  {author} {\bibinfo {author} {\bibfnamefont {R.}~\bibnamefont
			{Loudon}},\ }\href@noop {} {\emph {\bibinfo {title} {The quantum theory of
				light}}}\ (\bibinfo  {publisher} {OUP Oxford},\ \bibinfo {year}
	{2000})\BibitemShut {NoStop}%
\end{thebibliography}
%

\appendix

\section{Simplified Examples of Classical and Quantum Encryption}\label{appendix:examples}
Given the cross-disciplinary nature  of our work, it is prudent to provide a simplified introduction to quantum cryptography. Below, starting with a simple classical encryption using the (Vernam cipher) one-time pad (OTP) and quantum encryption through the BB84 QKD protocol, we discuss how a simple message may be encrypted using QKD keys via post-processing protocol. Other simplified QKD experiments and related theories include the work by Bloom et al.~\cite{physics4010009}. 

\subsection{Classical encryption}
As an example of classical encryption, suppose we would like to securely communicate the message ``dam’’ between an infrastructure point and a remote control unit. Let us write it  in binary representation based on ASCII values. In the ASCII standard, each character is represented using a unique 7-bit binary code. For instance, the binary representation for the character ``d'' is obtained by translating its ASCII representation (100) to a 7-bit binary string, yielding 1100100. Therefore,
\begin{equation}
\text{dam} \to 1100100\, 1100001\, 1101101.
\end{equation}
The one-time pad (OTP) is a symmetric-key encryption technique known to be unbreakable when used correctly. It applies a key (usually a random sequence of bits) to the message via a bitwise operation, typically, XOR to produce the ciphertext. To assure security using the OTP, the key should be as long as the plaintext, used only once, and kept secret. As our key, let us take the word ``key’’. The binary representation for ``key’’ is:
\begin{equation}
\text{key} \to 1101011\, 1100101\, 1111001.
\end{equation}
Let us now perform the XOR operation, which is defined as follows.
Given two binary values \( A \) and \( B \), the XOR (exclusive or) operation returns a value of 1 if the bits being compared are different, and 0 if they are the same. Formally:
\begin{equation}
A \oplus B = C,
\end{equation}
and, if we XOR the result \( C \) with \( B \), we retrieve the original value \( A \):
\begin{equation}
C \oplus B = A.
\end{equation}
Thus, for our ``$(\text{dam}\oplus \text{key})\to \text{ciphertext}$'' operation, we write:
\begin{align}
\text{1100100 (d)}\,\,\, & \oplus \,\,\,\text{1101011 (k)}\to 0001111,\\
\text{1100001 (a)}\,\,\, & \oplus \,\,\, \text{1100101 (e)}\to 0000100,\\
\text{1101101 (m)}\,\,\, & \oplus \,\,\, \text{1111001 (y)}\to 0010100.
\end{align}
Thus, the encrypted message is: 
\begin{equation}
\text{cyphertext:} \,\, 0001111 0000100 0010100. 
\end{equation}
To decipher the ciphertext produced using the OTP, we reapply the same XOR operation with the same key. This returns us to the original plaintext. Using our previously derived ciphertext and the key:
\begin{align}
\text{0001111}\,\,\, & \oplus \,\,\,\text{1101011 (k)}\to \text{1100100 (d)},\\
\text{0000100}\,\,\, & \oplus \,\,\, \text{1100101 (e)}\to \text{1100001 (a)},\\
\text{0010100}\,\,\, & \oplus \,\,\, \text{1111001 (y)}\to \text{1101101 (m)}.
\end{align}
Thus, when deciphered, the ciphertext using the key  gives us back the original message:
\begin{equation}
\text{plaintext:} \,\, 1100100\, 1100001\, 1101101 \to \text{``dam''}.
\end{equation}
This illustrates the reversible nature of the XOR operation: the encryption and decryption processes are effectively the same operation.
The generation of classical keys, especially for cryptographic purposes, is more intricate than just using a straightforward binary representation of a word, as we simplistically did above. The strength of cryptographic systems often hinges on the quality of the keys and the randomness or unpredictability of key generation. Classical keys may be generated via true, as well as pseudorandom random number generators. Key management practices, including generation, storage, distribution, rotation, and disposal, are vital.

\subsection{Quantum encryption}
The above encryption can also be performed using the key generated via QKD.  The communicating parties (Alice and Bob) could use a QKD protocol such as BB84 to generate a shared, secret random bit string. 
This process involves sending quantum states (e.g., photon polarizations) between Alice and Bob and performing post-processing following the steps in Fig.~\ref{fig:protocol} and Table~\ref{table:postp} (see
R. Wolf~\cite{ramona:book} for further reading). 
Alice can now use the shared key from the QKD process as the OTP key to XOR with her message, ``dam'' in binary.
\[ \text{dam} \rightarrow 1100100\ 1100001\ 1101101 \]
Suppose the QKD-generated key is \( K = k_1\ k_2\ k_3 \) (where each \( k_i \) is 7 bits in our toy example). She then XORs her message with this key to get the ciphertext:
\begin{align*}
1100100 \oplus k_1 &= c_1 \\
1100001 \oplus k_2 &= c_2 \\
1101101 \oplus k_3 &= c_3 
\end{align*}
The encrypted message is then \( c_1\ c_2\ c_3 \).
For decryption, Bob uses the same QKD-generated key to XOR with the received ciphertext to retrieve the original message.

\noindent \textbf{Step 1: Preparation \& Transmission by Alice}
\begin{itemize}
    \item Alice randomly selects bits and their corresponding bases. The bases can be:
    \begin{itemize}
        \item Rectilinear, represented as Z-basis: $\lvert 0 \rangle$ and $\lvert 1 \rangle$.
        \item Diagonal, represented as X-basis: $\lvert + \rangle = \frac{\lvert 0 \rangle + \lvert 1 \rangle}{\sqrt{2}}$ and $\lvert - \rangle = \frac{\lvert 0 \rangle - \lvert 1 \rangle}{\sqrt{2}}$.
    \end{itemize}
    \item For demonstration, consider the binary of "dam": 1100100 1100001 1101101. Alice selects the first 3 bits: 110.
    \begin{itemize}
        \item For the first bit (1), Alice chooses the rectilinear basis and sends $\lvert 1 \rangle$.
        \item For the second bit (1), Alice chooses the diagonal basis and sends $\lvert - \rangle$.
        \item For the third bit (0), Alice chooses the rectilinear basis and sends $\lvert 0 \rangle$.
    \end{itemize}
\end{itemize}

\noindent\textbf{Step 2: Measurement by Bob}
\begin{itemize}
    \item Bob randomly selects a basis for each received qubit and performs a measurement.
    \begin{itemize}
        \item For the first qubit, he chooses rectilinear and measures 1.
        \item For the second qubit, he chooses rectilinear (a mismatch with Alice) and gets a random result, say 0.
        \item For the third qubit, he chooses rectilinear and measures 0.
    \end{itemize}
\end{itemize}

\noindent\textbf{Step 3: Basis Discussion}
\begin{itemize}
    \item Alice and Bob publicly disclose the bases they used.
    \item They compare their choices:
    \begin{itemize}
        \item For the first bit, both chose rectilinear - they retain Bob's result.
        \item For the second bit, they used different bases - they discard Bob's result.
        \item For the third bit, both chose rectilinear - they retain Bob's result.
    \end{itemize}
    \item Their resulting raw key is now: 10.
\end{itemize}

\noindent\textbf{Step 4: Error Estimation}
\begin{itemize}
    \item A subset of the raw key is selected for error testing.
    \item They compare their respective bits in this subset publicly.
    \item Calculate QBER = (number of errors in subset)/(size of subset).
    \item If QBER exceeds a threshold, the protocol is aborted due to potential eavesdropping.
\end{itemize}

\noindent\textbf{Step 5: Privacy Amplification}
\begin{itemize}
    \item Aims to reduce any potential eavesdropper's information to an insignificant level.
    \item Two-universal hash functions might be applied to the key to produce a shorter, more secure key. For example, take the raw key to be: \(1011010101\). Consider a very basic hash function defined as:
    \begin{itemize}
        \item Break the string into groups of 2.
        \item For each group:
        \begin{itemize}
            \item If it is \(00\), it maps to \(0\).
            \item If it is \(01\) or \(10\), it maps to \(1\).
            \item If it is \(11\), it maps to \(0\).
        \end{itemize}
            Given the original key: $10 \quad 11 \quad 01 \quad 01 \quad 01,$ applying the hash function produces: $ 1 \quad 0 \quad 1 \quad 1 \quad 1,$
which is shorter than the original key as a result of a specific transformation.
    \end{itemize}
    \item Classical error-correcting codes can be used to rectify errors introduced by the quantum channel. For example, consider the Hamming(7,4) code~\cite{siehler2011hamming}:
    \begin{itemize}
        \item Designed to encode 4 bits of data into 7 bits by adding 3 parity bits.
        \item Given a 4-bit data `1101`, encoding adds parity bits to produce: 0 0 1 0 1 1 0.
        \item If an error flips the 6th bit during transmission, we receive: 0 0 1 0 1 0 0.
        \item The error is detected and corrected using the parity bits, restoring the original encoded string.
    \end{itemize}
\end{itemize}
Our plaintext message ``dam’’ has a length of 21 bits, as noted above. Due to the probabilistic nature of quantum measurements and the random choice of bases, typically only around 50\% of the initially sent qubits contribute to the raw key post key sifting. So, if Alice wants to ensure a shared secret key of length 21 bits (to match the plaintext message length), she will need to initiate the process with more than 42 qubits (assuming a 50\% retention rate after sifting). Thus, the process requires the transmission of a greater number of qubits than the intended message length due to the key sifting process and potential eavesdropping checks.

\section{Tables}
\begin{table}[H]
    \centering
    \caption{Brief List of Basic Cybersecurity Issues and the Role of QKD for Dams}\label{table}
    \begin{tabularx}{\textwidth}{X|X|X|X}
        \toprule
        \textbf{Cybersecurity Issue} & \textbf{Description} & \textbf{Impact} & \textbf{Role of QKD} \\ 
        \midrule
        Remote Control System Attacks & Compromise of SCADA~\cite{Alrefaei2022AnOO} systems controlling dam operations. & Dam failure, potential loss of life, and environmental damage. & Secure keys via QKD encrypt communication, thwarting unauthorized access to the control system. \\
        \midrule
        Sensor Spoofing~\cite{urbina2016survey} & Interference with sensors, leading to inaccurate readings and unsafe operations. & Dam failure, potential loss of life, and environmental damage. & Secure or authenticated  communication between sensors and control system, reveals data tampering. \\
        \midrule
        Communication Interception~\cite{chen2011lessons} & Interception or injection of malicious commands in communication channels. & Dam failure, potential loss of life, and environmental damage. & Secure all communications, hindering interception or data tampering. \\
        \midrule
        Denial of Service (DoS) Attacks~\cite{lee2016analysis} & Overloading communication channels or control systems. & Operational disruption leading to flooding or other issues. & Indirectly aids by protecting from vulnerabilities exploited in DoS. \\
        \midrule
        Physical Security Breaches & Tampering of equipment or insertion of malicious hardware/software. & Dam failure, potential loss of life, and environmental damage. & Indirectly aids by strengthening overall cybersecurity infrastructure. \\
        \midrule
        Supply Chain Attacks & Pre-installation compromise of hardware or software. & Compromised dam components leading to security breaches. & Indirectly aids by protecting from vulnerabilities exploited due to compromised components. \\
        \midrule
        Insider Threats & Misuse of sensitive systems by authorized individuals. & Operational disruption or sabotage. & Secure communication between control systems and authorized personnel, preventing unauthorized access. \\
        \bottomrule
    \end{tabularx}
\end{table}
\addtocounter{table}{-1}
\begin{table}[H]
    \centering
    \caption{Typical Noise Sources in a Dam Environment and Corresponding Sensors with Quantitative Descriptions (refer to Ouellet et al.~\cite{ouellet2022advanced} for monitoring relevant noise sources, and other works~\cite{geosciences9040187,petr,MOHANTA2017637,quaranta2021noise} for specific example of noise frequency and amplitude.)}
    \label{noise}
    \begin{tabularx}{\textwidth}{>{\hsize=0.9\hsize}X|>{\hsize=1.2\hsize}X|>{\hsize=0.9\hsize}X}
        \toprule
        \textbf{Noise Source} & \textbf{Description} & \textbf{Typical Sensor} \\
        \midrule
        Turbine Operations & Noise from turbine movement both in air and underwater. Frequencies $f \sim (0.5-30)$~Hz with amplitudes $\lesssim 1$~mm. & Hydrophone (underwater), Microphone (airborne) \\
        \midrule
        Gates and Valves & Noise due to dam gate or spillway operations. Varies based on size and operation speed. & Vibration sensors, Microphone \\
        \midrule
        Pumps and Machinery & Noise from operational machinery. Typically $f \sim (10-200)$~Hz. & Microphone, Vibration sensors \\
        \midrule
        Flow Turbulence & Noise from rapid and turbulent water flow. $f \sim (1-100)$~Hz. & Hydrophone \\
        \midrule
        Waterfall/Spill & Noise due to water spillage. Frequency depends on water volume and height of fall. & Hydrophone, Microphone \\
        \midrule
        Bubble Formation & Noise due to bubble formation and collapse. $f \sim (5-50)$~Hz. & Hydrophone \\
        \midrule
        Transformer Operations & Buzzing or humming from transformers. Typically at 50 Hz or 60 Hz. & Magnetic field sensors, Microphone \\
        \midrule
        High Voltage Equipment & Noise from insulator discharges. Broadband noise typically spanning 10 Hz to 1 kHz. & Electromagnetic sensors, Microphone \\
        \midrule
        Vibration & Vibrations inherent to dam structures. Spanning from very low frequencies (<1 Hz) due to seismic activities to high frequencies (>100 Hz) from machinery operations. & Accelerometers, Vibration sensors \\
        \midrule
        Thermal Expansion/Contraction & Noise from temperature-induced structural changes. Frequency varies based on structure size and material. & Vibration sensors, Microphone \\
        \midrule
        Wildlife Activities & Sounds from local fauna. Frequencies are species-specific, ranging broadly from 1 Hz to 10 kHz. & Microphone, Hydrophone \\
        \midrule
        Weather Patterns & Noise from atmospheric disturbances, thunder, tornado. Broad frequency range from <1 Hz (thunder rumble) to >10 kHz (lightning crack). & Wind sensors, Microphone \\
        \midrule
        Vehicle Traffic & Noise from vehicular activities. Frequencies range from 20 Hz (engine hum) to 2 kHz (horn). & Accelerometers, Vibration sensors, Microphone \\
        \midrule
        Construction/Maintenance & Noise from maintenance or construction work. Broad frequency range depending on tools and machinery. & Accelerometers, Microphone, Vibration sensors \\
        \midrule
        Temperature Fluctuations & Ambient temperature changes affecting equipment. Changes can cause material contractions or expansions leading to noise. & Thermocouples, Infrared sensors \\
        \midrule
        Moisture/Condensation & Moisture interference with equipment. Can cause electrical noises or material deformations. & Humidity sensors, Moisture meters \\
        \bottomrule
    \end{tabularx}
\end{table}
\addtocounter{table}{-1}
\begin{table}[H]
    \centering
    \caption{Basic Noise Sources in QKD Systems in Laboratory Settings (see also \cite{scarani2009security,gobby2004quantum,yuan2007high,rosenberg2005practical,hiskett2007long})}
    \label{intrinsic:noise}
    \begin{tabularx}{\textwidth}{l|X}
        \toprule
        \textbf{Noise Source} & \textbf{Description} \\
        \midrule
        Quantum Bit Error Ratio (QBER) & 
        Represents the ratio of bits that are received in error. Though not a direct noise source, QBER quantifies the impact of various technical factors and imperfections in QKD systems. \\
        \midrule
        Dark Counts & 
        False counts arising in photon detectors due to thermal fluctuations or other non-signal measurement events. \\
        \midrule
        Dead Time & 
        Time taken by a detector to recover after detecting a photon. Photons arriving during this interval can lead to loss. \\
        \midrule
        Detector Jitter & 
        Uncertainty in a detector's time response when it receives a signal, arising from electronic and photonic fluctuations. \\
        \midrule
        Beam Splitting/Coupling Inefficiencies & 
        Imperfections in beam splitters or inefficient coupling into optical fibers leading to photon loss. \\
        \midrule
        Fiber or Channel Attenuation & 
        Losses in the optical channel or the transmission fiber. \\
        \midrule
        Multi-Photon Emissions & 
        Occurrences when sources produce multi-photon pulses, introducing vulnerabilities and noise. \\
        \midrule
        Phase Fluctuations & 
        In protocols like Differential Phase Shift QKD, phase fluctuations in transmission fiber can cause errors. \\
        \midrule
        Timing Jitter/Synchronization & 
        Uncertainty or variations in the timing of a system's clock or reference signal, affecting synchronization. \\
        \midrule
        Quantum State Preparation & 
        Imperfections in preparing quantum states like specific polarization states. \\
        \midrule
        Spatial Mode Mismatches & 
        Mismatches when transmitting quantum states over channels, leading to decreased detection probabilities. \\
        \midrule
        Back Reflections/Scattering & 
        Reflections from interfaces or scattering within components introducing noise photons. \\
        \bottomrule
    \end{tabularx}
\end{table}

\addtocounter{table}{-1}
\begin{table}[H]
    \centering
    \caption{Comparison of QKD Modalities (see e.g., Diamanti et al.~\cite{diamanti2016practical}) against Hydropower System's Requirements}
    \begin{tabularx}{\textwidth}{| p{4cm} | X |}
        \toprule
        \textbf{QKD Modality} & \textbf{Hydropower System's Requirements} \\
        \midrule
        BB84 Protocol (see e.g. Nadal et al.~\cite{nandal2020survey} for other protocols) & 
        \begin{itemize}
            \item Suitable for existing optical fiber networks.
            \item Direct point-to-point setups for small-scale plants.
            \item Economically efficient for short distances.
        \end{itemize} \\
        \midrule
        Decoy State QKD & 
        \begin{itemize}
            \item Robust against photon number splitting attacks.
            \item Beneficial for medium to large-scale plants with potential eavesdropping threats.
        \end{itemize} \\
        \midrule
        Continuous-Variable QKD & 
        \begin{itemize}
            \item Preferred for metropolitan-area networks.
            \item Requires direct trusted relay and quantum repeaters for long distances.
            \item Suitable for high transmission rate requirements.
        \end{itemize} \\
        \midrule
        MDI-QKD & 
        \begin{itemize}
            \item Ideal for infrastructures at risk from sophisticated adversaries.
            \item Eliminates detector side-channel vulnerabilities at center detection node.
            \item May require higher initial investment for equipment.
        \end{itemize} \\
        \midrule
        Satellite-based QKD & 
        \begin{itemize}
            \item Best for remote facilities over vast areas.
            \item Capital-intensive but offers broad coverage.
            \item Enables global scale secure communications but at low rates.
            \item Requires clear sky conditions for optimal operations.
        \end{itemize} \\
        \midrule
        Other Considerations &
        \begin{itemize}
            \item Maintenance and operational costs.
            \item Scalability to future expansions.
            \item Interoperability with existing communication systems.
            \item Training and expertise requirements.
            \item Key management/revocation/lawful intercept requirements
        \end{itemize} \\
        \bottomrule
    \end{tabularx}
    \label{tab:qkd_comparison}
\end{table}
\addtocounter{table}{-1}
\begin{table}[H]
    \centering
    \caption{Basic Post-processing Steps in QKD (see Fig.~\ref{fig:protocol}, and R. Wolf~\cite{ramona:book} for further reading)}
    \label{table:postp}
    \begin{tabularx}{\textwidth}{l|X}
        \toprule
        \textbf{Step} & \textbf{Description/Equation} \\
        \midrule
        Error Estimation & 
        \( QBER = \frac{\text{number of error bits}}{\text{total number of bits exchanged}} \) \\
        \midrule
        Information Reconciliation & 
        Uses error-correcting codes to rectify key discrepancies. The Cascade protocol is popular; it entails key division, shuffling, and parity comparison. \\
        \midrule
        Privacy Amplification & 
        Aims to eliminate any eavesdropper's partial information. Typically employs universal hash functions, represented as: \( k = f(k_{\text{raw}}) \). \\
        \midrule
        Key Sifting & 
        Particularly relevant in the BB84 protocol. Alice and Bob publicly disclose the bases chosen for each qubit. Qubits with differing bases are discarded. \\
        \midrule
        Authentication & 
        Confirms genuine communication between Alice and Bob. Utilizes classical authentication methods in tandem with previously shared secret keys. \\
        \bottomrule
    \end{tabularx}
\end{table}
\addtocounter{table}{-1}

\section{Glossary}
\begin{tabularx}{\textwidth}{>{\hsize=.5\hsize}X|>{\hsize=1.5\hsize}X}
    \caption{Glossary of Terms and Definitions}
    \label{table:glossary}
    \\
    \toprule
    \textbf{Term} & \textbf{Definition} \\
        \midrule
        $a$ & Fiber's Attenuation: Represented in units of dB/km, it is the property of the optical fiber that quantifies the loss of signal strength per unit length of fiber. \\
        \midrule
        Alice and Bob & Conventionally used names to denote the sender and receiver in cryptographic communications, including in QKD systems. \\
        \midrule
        Attenuation & The reduction of signal strength as it travels through a medium, such as an optical fiber, due to absorption, scattering, and other loss mechanisms. \\
        \midrule
        BB84 & A quantum key distribution protocol developed in 1984 by Bennett and Brassard, using two non-orthogonal bases~\cite{wilde_2017}. \\
        \midrule
        Binary Entropy Function & A function quantifying the maximum possible information about the total key based on shared bits between parties. Typically denoted as \( H(p) \), it represents the uncertainty of a binary random variable and is defined as~\cite{ramona:book,wilde_2017}:
        $$H(p) = -p \log_2(p) - (1-p) \log_2(1-p),$$ 
        where \( p \) is the probability of one of the two outcomes (e.g., a bit being 1). Consequently, \( 1-p \) is the probability of the other outcome (the bit being 0). $H$ is defined for \( 0 \leq p \leq 1 \) with a maximum of 1, which occurs when \( p = 0.5 \), indicating maximum uncertainty (i.e., both outcomes are equally probable). When \( p = 0 \) or \( p = 1 \), \( H(p) = 0 \), there is no uncertainty. In our work, Eq.~\ref{eq2} yields the error rate in a specific basis ($x$ basis) and represents the maximum possible information rate that can be deduced about the total key based on shared bits for error estimation. \\
        \midrule
        Channel Loss Parameter ($\eta_{ch}$) & A dimensionless parameter derived from the total optical loss ($\gamma$) and representing the linear loss due to the fiber optic cables in the QKD system. \\
        \midrule
        Coherent States & A coherent quantum state \( | \alpha \rangle \) is defined as:
        $$ | \alpha \rangle = e^{-\frac{|\alpha|^2}{2}} \sum_{n=0}^{\infty} \frac{\alpha^n}{\sqrt{n!}} | n \rangle,$$
        where \( \alpha \) is a complex number and \( | n \rangle \) are the photon number (Fock) states~\cite{loudon2000quantum}. Such states play a fundamental role in quantum optics due to their semi-classical nature and are crucial in various quantum communication protocols, including QKD~\cite{ramona:book}.\\
        \midrule
        Continuous Variable-Quantum Key Distribution (CV-QKD) & CV-QKD employs continuous quantum variables such as the quadratures of the electromagnetic field to encode information. The most common CV-QKD protocols are based on coherent states using Gaussian modulation of amplitude and phase, and they use homodyne or heterodyne detection for the decoding process. This approach has the advantage of being compatible with conventional telecom technology, potentially allowing for more straightforward integration into existing networks. However, it typically makes more assumptions that detector noise can be ``trusted,'' and calibrated away. \\
        \midrule
        Dark Count & Counts detected in the absence of light, usually due to noise or unwanted signals. \\
        \midrule
        Decoy State QKD & A variation of the BB84 QKD protocol which employs additional signal states to improve security against photon number splitting attacks. \\
        \midrule
        Discrete Variable (DV) QKD & Discrete Variable (DV) QKD leverages the quantum properties of individual photons to secure communication. It typically operates using polarization or phase encoding schemes to encode the quantum bits (qubits). Protocols such as BB84, B92, and SARG04 are well-known in the DV-QKD sphere~\cite{nandal2020survey,kong2020review}. \\
        \midrule
        Error Correction & A process to identify and correct errors in the quantum key transmission. \\
        \midrule
        Error Rate in the X Basis ($\phi_x$) & A parameter representing the rate of error in the x basis of the key during QKD operations. \\
        \midrule
        Error Reconciliation & A procedure in QKD to correct any discrepancies in the key between the two parties. \\
        \midrule
        Eve & Eavesdropper: Conventionally used name to denote a potential attacker trying to gain unauthorized access to the secure communication. All errors are typically attributed to her.\\
        \midrule
        Fiber-Based Loss & Refers to the loss of signal in optical fibers, affecting the transmission of quantum signals. \\
        \midrule
        Fiber Length ($L$) & The physical distance covered by the optical fiber, usually represented in kilometers. \\
        \midrule
        Finite Size Effects & Describes the effects or limitations of having a finite number of signals in QKD. \\
        \midrule
        Homodyne Detection & A technique used in quantum cryptography for measuring a quantum signal through interference with a local oscillator using balanced difference detection. \\
        \midrule
        Hydropower & The generation of power through the use of the gravitational force of falling or fast-running water. \\
        \midrule
        Information Theoretical Security (ITS) & A security paradigm that assures confidentiality regardless of the computational resources of an adversary. \\
        \midrule
        Leakage During Error Correction ($\text{Leak}_{EC}$) & The segment of key information that might be exposed to an eavesdropper during error correction procedures in the QKD protocol. \\
        \midrule
        Misalignment Angle & Refers to the variation in the angle of the initial polarization state, which can be caused by factors such as thermal fluctuations or physical stress on the fiber. \\
        \midrule
        One-Time Pad (OTP) & A method of encryption where a message is combined with a key using exclusive OR (XOR) operation. \\
        \midrule
        Phase Modulation & The modulation of the phase of a carrier signal to encode information, often used in QKD systems to encode quantum information. \\
        \midrule
        Photon Avalanche Detectors & Photodetectors that can detect low-intensity light down to single photons, often used in QKD setups. \\
        \midrule
        Polarization & Refers to the orientation of oscillations in electromagnetic waves, used to encode information in quantum states in the context of QKD. \\
        \midrule
        Quantum Bit Error Rate (QBER) & The rate of errors that occur during quantum transmission. \\
        \midrule
        Quantum Cryptography (QC) & A method of securing communication channels by applying quantum mechanics principles. \\
        \midrule
        Quantum Key Distribution (QKD) & A cryptographic protocol based on quantum mechanics to securely distribute random private keys between two parties. \\
        \midrule
        Secure Key Length ($L$) & The length of the secure key, here denoted $L_{\text{key}}$, in a QKD system, chosen based on various parameters to maintain a balance between security and system performance. \\
        \midrule
        Secure Key Rate (SKR) & The rate at which a QKD system can produce secure shared private keys, influenced by factors like distance and error rate. A metric to evaluate the performance of a QKD system\\
        \midrule
        Secure Rate Formula & A mathematical representation of the rate at which a QKD system can generate secure keys. \\
        \midrule
        Total Optical Loss ($\gamma$) & Represented in units of dB, it measures the total loss in the system, arising due to the fiber's attenuation ($a$) and the length of the fiber ($L$). \\
    \bottomrule
\end{tabularx}




\end{document}